\documentclass[10pt,aps,prb,twocolumn, nofootinbib]{revtex4-1}

\usepackage{amssymb,amsmath}
\usepackage{subfig}
\usepackage{graphicx}
\usepackage{color}
\usepackage{hyperref}
\usepackage{listings}
\usepackage{caption}
\usepackage{changepage}
\hypersetup{
    unicode=false,          % non-Latin characters in Acrobat’s bookmarks
    pdftoolbar=true,        % show Acrobat’s toolbar?
    pdfmenubar=true,        % show Acrobat’s menu?
    pdffitwindow=false,     % window fit to page when opened
    pdfstartview={FitH},    % fits the width of the page to the window
    pdfauthor={R.S. van Bergen & N. Kriegeskorte},     % au\textit{}thor
    colorlinks=true,       % false: boxed links; true: colored links
    linkcolor=blue,          % color of internal links
    citecolor=red,        % color of links to bibliography
}
\graphicspath{{figures/}}
\usepackage{natbib}
\def\url#1{}

\begin{document}

\definecolor{dkgreen}{rgb}{0,0.6,0}
\definecolor{gray}{rgb}{0.5,0.5,0.5}
\definecolor{mauve}{rgb}{0.58,0,0.82}

\title{\huge Going in circles is the way forward:\\ \LARGE the role of recurrence in visual inference}
\author{Ruben S. van Bergen$^1$, Nikolaus Kriegeskorte$^{1-4}$}
\affiliation{$^1$Zuckerman Mind Brain Behavior Institute, Columbia University, New York, NY, United States}
\affiliation{$^2$Department of Psychology, Columbia University, New York, NY, United States}
\affiliation{$^3$Department of Neuroscience, Columbia University, New York, NY, United States}
\affiliation{$^4$Affiliated member, Electrical Engineering, Columbia University, New York, NY, United States}

\begin{abstract}
\noindent Biological visual systems exhibit abundant recurrent connectivity. State-of-the-art neural network models for visual recognition, by contrast, rely heavily or exclusively on feedforward computation. Any finite-time recurrent neural network (RNN) can be unrolled along time to yield an equivalent feedforward neural network (FNN). This important insight suggests that computational neuroscientists may not need to engage recurrent computation, and that computer-vision engineers may be limiting themselves to a special case of FNN if they build recurrent models. Here we argue, to the contrary, that FNNs are a special case of RNNs and that computational neuroscientists and engineers should engage recurrence to understand how brains and machines can (1) achieve greater and more flexible computational depth, (2) compress complex computations into limited hardware, (3) integrate priors and priorities into visual inference through expectation and attention, (4) exploit sequential dependencies in their data for better inference and prediction, and (5) leverage the power of iterative computation. 
\end{abstract}

\maketitle

\sffamily{}

\section*{\label{sec:one}Introduction} \noindent
The primate visual cortex uses a recurrent algorithm to process sensory input\cite{Lamme2000,Kreiman,Angelucci2006}. Anatomically, connectivity is cyclic. Neurons are connected in cycles within local cortical circuits \cite{Anderson1994,Martin2002,Douglas2007}. Global inter-area connections are dense and mostly bidirectional \cite{Felleman1991,Salin1995,Markov2014}. Physiologically, the dynamics of neural responses bear temporal signatures indicative of recurrent processing \cite{Douglas1995,Lamme2000,Super2001}. Behaviorally, visual perception can be disturbed by carefully timed interventions that coincide with the arrival of re-entrant information to a visual area \cite{DiLollo2000,Lamme2001,Heinen2005,Fahrenfort2007}. The evidence for recurrent computation in the primate brain, thus, is unequivocal. What is less obvious, however, is \textit{why} the brain uses a recurrent algorithm. 
\begin{figure}[b!]
    %\centering
    \includegraphics[width=0.45\textwidth]{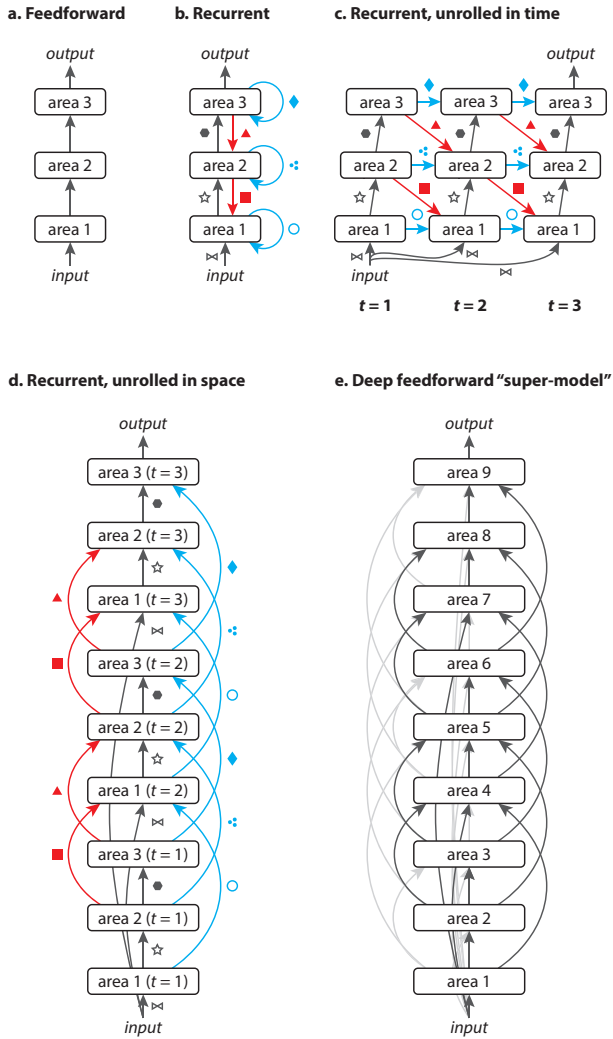}
    \caption{\textbf{Unrolling recurrent neural networks.} (\textbf{a}) A simple feedforward neural network. (\textbf{b}) The same network with lateral (blue) and feedback (red) connections added, to make it recurrent. (\textbf{c}) "Unrolling" the network in time clarifies the order of its computations. Here, the network is unrolled for three time steps before its output is read out, but we could choose to run the network for more or fewer steps. Areas are staggered from left to right to show the order in which their neural activities are updated. (\textbf{d}) Alternatively, we can unroll the recurrent network's time steps in space, by arranging the areas and connections from different time steps in a linear spatial sequence. Note how all arrows now once again point in the same (forward) direction, from input to output. Throughout panels (\textbf{a}-\textbf{b}), connections that are identical (sharing the same weight matrices) are indicated by corresponding symbols. (\textbf{e}) If we lift the weight-sharing constraints from the previous network, this induces a deep feedforward "super-model", which can implement the spatially-unrolled recurrent network as a special case. This more general architecture may include additional connections (examples shown as light gray arrows) not present in the spatially-unrolled recurrent net.}
    \label{fig:unroll}
\end{figure}

This question has recently been brought into sharper focus by the successes of deep feedforward neural network models (FNNs)\cite{Lecun2015,Schmidhuber2015}. These models now match or exceed human performance on certain visual tasks \cite{He2015,He2016,Kemelmacher-Shlizerman2016}, and better predict primate recognition behavior \cite{Kubilius2016,Majaj2018,Spoerer2019} and neural activity \cite{Cadieu2014,Khaligh-Razavi2014,Guclu2015,Kriegeskorte2015,Kheradpisheh2016,Schrimpf2018} than current alternative models.

Although computer vision and computational neuroscience both have a long history of recurrent models\cite{Rao1999,Yuille2006,Friston2009,Prince2012}, feedforward models have earned a dominant status in both fields. How should we account for this discrepancy between brains and models? 

One answer is that the discrepancy reflects the fact that brains and computer-vision systems operate on different hardware and under different constraints on space, time, and energy. Perhaps we have come to a point at which the two fields must go their separate ways. However, this answer is unsatisfying. Computational neuroscience must still find out how visual inference works in brains. And although engineers face \textit{quantitatively} different constraints when building computer-vision systems, they, too, must care about the spatial, temporal, and energetic limitations their models must operate under when deployed in, for example, a smartphone. Moreover, as long as neural network models continue to dominate computer vision, more efficient hardware implementations are likely to be more similar to biological neural networks than current implementations using conventional processors and graphics processing units (GPUs).

A second explanation for the discrepancy is that the abundance of recurrent connections in cortex belies a superficial role in neural computation. Perhaps the core computations can be performed by a feedforward network \cite{DiCarlo2012}, while recurrent processing serves more auxiliary and modulatory functions, such as divisive normalization\cite{Carandini2012} and attention\cite{Desimone1995,Kastner2000,Maunsell2006,NIPS2017_7181}. This perspective is convenient because it enables us to hold on to the feedforward model in our minds. The auxiliary and modulatory functions let us acknowledge recurrence without fundamentally changing the way we envision the algorithm of recognition.

However, there is a third and more exciting explanation for the discrepancy between recurrent brains and feedforward models: Although feedforward computation is powerful, a recurrent algorithm provides a fundamentally superior solution to the problem of visual inference, and this algorithm is implemented in primate visual cortex. This recurrent algorithm explains how primate vision can be so efficient in terms of space, time, energy, and data, while being so rich and robust in terms of the inferences and their generalization to novel environments.

In this review, we argue for the latter possibility, discussing a range of potential computational functions of recurrence and citing the evidence suggesting that the primate brain employs them. We aim to distinguish established from more speculative, and superficial from more profound forms of recurrence, so as to clarify the most exciting directions for future research that will close the gap between models and brains.

\section*{\label{sec:two}Unrolling a recurrent network} \noindent
What exactly do we mean when we say that a neural network -- whether biological or artificial -- is recurrent rather than feedforward? This may seem obvious, but it turns out that the distinction can easily be blurred. Consider the simple network in \textbf{Fig. \ref{fig:unroll}a}. It consists of three processing stages, arranged hierarchically, which we will refer to as \textit{areas}, by analogy to cortex. Each area contains a number of neurons (real or artificial) that apply fixed operations to their input. Visual input enters in the first area, where it undergoes some transformation, the result of which is passed as input to the second area, and so forth. Information travels exclusively in one direction -- the “forward” direction, from input to output -- and so this is an example of a feedforward architecture. Notably, the number of transformations between input and output is fixed, and equal to the number of areas in the network. 

\begin{figure*}
    %\centering
    \includegraphics[width=0.7\textwidth]{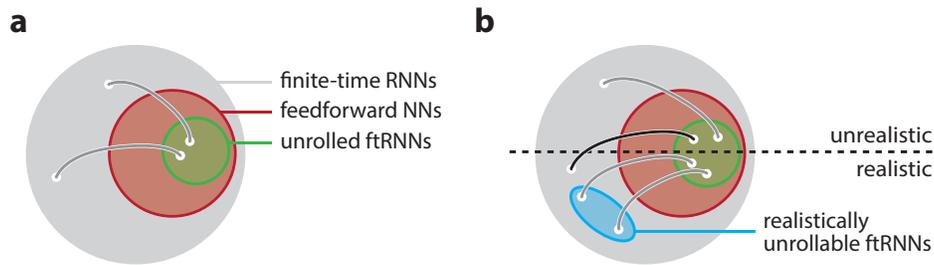}
    \caption{\textbf{Relationships between recurrent and feedforward networks.} This figure illustrates relationships between discrete-time feedforward (FNN) and discrete-time recurrent (RNN) neural network models. (\textbf{a}) The architecture of any RNN can be reduced to an FNN by removing all its recurrent connections (e.g., going from \textbf{Fig. \ref{fig:unroll}b} back to \textbf{Fig. \ref{fig:unroll}a}), or equivalently, setting the weights of these connections to zero. Vice versa, any FNN can be expanded to an infinite variety of RNNs by adding lateral or feedback connections. Feedforward networks, thus, form an architectural subset of RNNs. Here we specifically consider RNNs that accomplish their task in a finite number of time steps. These finite-time RNNs (ftRNNs) have the special property that they can be unrolled into equivalent FNNs. White points linked by arcs indicate pairs of computationally equivalent architectures. Thus, the feedforward NNs contain a subset of architectures that can be obtained by unrolling a ftRNN. (\textbf{b}) These sets of networks can be further subdivided into subsets that are or are not \textit{realistic} to implement with the computational resources available for a brain or engineered device (areas below and above the dotted line, respectively). Deeper networks and, more generally, networks with more neurons and connections tend to require more memory and computation to train and run. Some realistic ftRNNs remain realistic when expressed as an FNN (blue ellipse). Others, however, become too complex, when unrolled, to be feasible (black arc crossing the realism line). This is because the unrolling operation induces a much deeper architecture with many more neural connections to be stored. These not-realistically-unrollable ftRNNs are especially interesting, since they correspond to recurrent solutions that cannot be replaced by feedforward architectures.}
    \label{fig:setmap}
\end{figure*}

Now compare this to the architecture in \textbf{Fig. \ref{fig:unroll}b}. Here, we have added lateral and feedback connections to the network. Lateral connections allow the output of an area to be fed back into the same area, to influence its computations in the next processing step. Feedback connections allow the output of an area to influence information processing in a lower area. There is some freedom in the order in which computations may occur in such a network. The order we illustrate here starts with a full feed-forward pass through the network. In subsequent time steps, neural activations are updated in ascending order through the hierarchy, based on the activations that were computed in the previous time step. 

This order of operations can be seen more clearly if we 'unroll' the network in time, as shown in \textbf{Fig. \ref{fig:unroll}c}. In this illustration, the network is unrolled for a fixed number of time steps (3). In fact, recurrent processing can be run for any desired duration before its output is read out -- a notion we will return to later. Notice how this temporally unrolled, small network resembles a larger feedforward neural network with more connections and areas between its input and output. We can emphasize this recurrent-feedforward equivalence by interpreting the computational graph over time as a spatial architecture, and visually arranging the induced areas and connections in a linear spatial sequence -- an operation we call \textit{unrolling in space} (\textbf{Fig. \ref{fig:unroll}d}). This results in a deep feedforward architecture with many \textit{skip connections} between areas that are separated by more than one level in this new hierarchy, and with many connections that are exact copies of one another (sharing identical connection weights). 

Thus, any finite-time RNN can be transformed into an equivalent FNN. But this should not be taken to mean that RNNs are a special case of FNNs. In fact, FNNs are a special case of finite-time RNNs (\textbf{Fig. \ref{fig:setmap}a}), comprising those which happen to have no cycles. More practically, not every unrolled finite-time RNN is a \textit{realistic} FNN (\textbf{Fig. \ref{fig:setmap}b}). By realistic networks, we mean networks that conform to the real-world constraints the system must operate
under. For computational neuroscience, a realistic network is one that fits in the brain of the animal and does not require a deeper network architecture or more processing steps than the animal can accommodate. For computer vision, a realistic network is one that can be trained and deployed on available hardware at the training and deployment stages. For example, there may be limits on the storage and energy available, which would limit the complexity of the architecture and computational graph. A realistic finite-time RNN, when unrolled, can yield an unworkably deep FNN. Although the most widely used current method for training RNNs (backpropagation through time) requires unrolling, an RNN is not equivalent to its unrolled FNN twin at the stage of real-world deployment: the RNN's recurrent connections need not be physically duplicated, but can be reused across cycles of computation.

An important recent observation\cite{Liao2016,Jastrzebski2017,Greff2019} is that the architecture that results from spatially unrolling a recurrent network, resembles the architectures of state-of-the art FNNs used in computer vision, which similarly contain skip connections and can be very deep. These deep FNNs may form a super-class of models (\textbf{Fig. \ref{fig:unroll}e}), which reduce to “recurrent-equivalent” architectures when certain subsets of weights are constrained to be identical. Liao \& Poggio \cite{Liao2016} showed that deep feedforward architectures known as \textit{residual networks} (ResNets) \cite{He2016} are formally equivalent to recurrent architectures when certain connection weights are constrained to be identical. Moreover, when ResNets were trained with such recurrent-equivalent weight-sharing constraints, their performance on computer vision benchmarks was similar to unconstrained ResNets (even though the weight sharing drastically reduces the parameter count and limits the component computations that the network can perform). This is especially noteworthy given that ResNets, and architecturally related DenseNets, are currently among the top-ranking FNNs on prominent computer vision benchmarks \cite{He2016,Huang2017}, as well as measures of brain-similarity \cite{Schrimpf2018}. Today's best artificial vision models, thus, actually implement computational graphs closely related to those of recurrent networks, even though these models are strictly feedforward architectures.

\section*{Continuous- versus discrete-time dynamics} \noindent
FNNs used in computer vision do not have meaningful dynamics. Each unit in the network instantaneously transforms its input into an output. This is in contrast to a feedforward network of biological neurons. When given a static input, biological neurons do not immediately produce their final responses. The movement of electric charges and neurotransmitters, and the opening and closing of ion channels takes time, so the network will gradually transition from its initial to its final state, with its trajectory continually perturbed by noise. Such continuous-time dynamics can be described by differential equations. When these cannot be solved analytically (as is typically the case), the dynamics can be simulated in discrete steps. In each step, the current state of each simulated neuron is updated. The future state of the network thus depends on its current state, as it does in an RNN. Consequently, the computational graph of the simulation algorithm contains loops from each neuron back to itself. Running the simulation over time amounts to unrolling this loopy computational graph, even though the network architecture did not contain loops.

Computational neuroscientists commonly study models of feedforward and recurrent neural networks with continuous-time dynamics\cite{Dayan2001}. Here our focus is on neural network models that are motivated by the goal to capture computations, rather than their precise neural implementation. The discrete-time behavior of such a model is not derived from a continuous-time description in differential equations. Moreover, the model is optimized in its discrete-time implementation. However, an implicit assumption in the field is that such models could be implemented in biological brains, and thus in continuous-time dynamical systems.

\section*{\label{sec:three}Reasons to recur}  \noindent
We have described how a recurrent network can be unrolled into a deep feedforward architecture. The resulting feedforward super-model offers greater computational flexibility, since weight-sharing constraints can be omitted and additional skip connections added to the network (\textbf{Fig. \ref{fig:unroll}e}). So what would be the benefit of restricting ourselves to recurrent architectures? We will first discuss the benefits of recurrence in terms of overarching principles, before considering more specific implementations of these principles.

\subsection*{Recurrence provides greater and more flexible computational depth}

\subsubsection*{Recurrence enables arbitrary computational depth} \noindent
One important advantage of recurrent algorithms is that they can be run for any desired length of time before their output is collected. We can define computational depth as the maximum path length (i.e. number of successive connections and nonlinear transformations) between input and output. A recurrent neural network (RNN) can achieve arbitrary computational depth despite having a finite count of parameters and being limited to finite spatial components. In other words, it can multiply its limited spatial resources along time. These deeper computations can serve to expand on the number of hypotheses considered (in generative inference) or on the number of   nonlinear features computed (in discriminative inference), or to extend the representation into the future or past, or to iteratively converge to a good estimate of some latent variable of interest.

\subsubsection*{Recurrence enables more flexible expenditure of energy and time in exchange for inferential accuracy} \noindent
In addition to enabling an arbitrarily deep computation given enough time, an RNN can \textit{adjust} its computational depth to the task at hand. The computational depth of a feedforward net, by contrast, is a fixed number determined by the architecture.

\begin{figure}[!b]
    %\centering
    \includegraphics[width=0.475\textwidth]{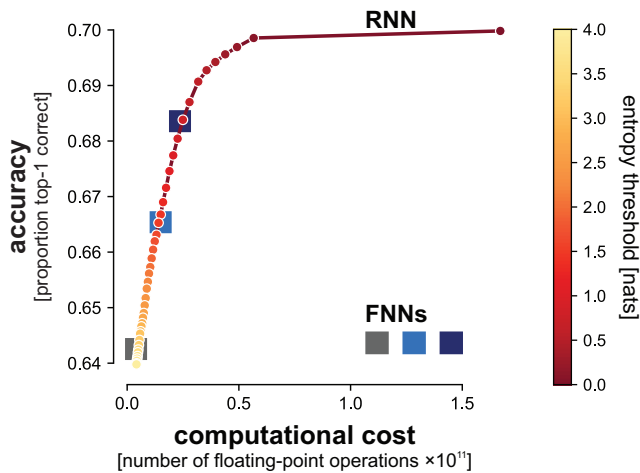}
    \caption{\textbf{Recurrence enables a network to trade speed for accuracy while approximately emulating the accuracies of feedforward models on average at matched computational cost.} Circles denote the performance of a recurrent neural network (RNN) that was run for different numbers of time steps, until it achieved a desired threshold of image classification confidence (quantified by the entropy of the class probabilities in the final network layer). Squares correspond to three architecturally similar feedforward networks (FNN) with different computational costs. On the x-axis is the computational cost of running these models, measured by the number of floating point operations. For the feedforward models, this cost is fixed by the architecture. For the recurrent models, it is the average number of operations that was required to meet the given entropy threshold. The y-axis shows the classification accuracy achieved by each model.  The performance of the recurrent model for different certainty thresholds follows a smooth curve, trading off computational cost (and thus computational speed) and accuracy. Note that this curve passes almost exactly through the cost-accuracy combinations achieved by the feedforward models. Thus, a single recurrent model can emulate the performance of multiple feedforward models as it trades off speed and accuracy. When the confidence threshold of termination was set such that the RNN matched the accuracy of a given FNN, the RNN required a similar number of floating-point operations on average as the FNN.
    (Figure adapted with permission from the authors\cite{Spoerer2019}.)
    }
    \label{fig:speedacc}
\end{figure}

Spoerer et al. implemented a recurrent model that terminates computations when it reaches a confidence threshold (defined by the entropy of the posterior, a measure of the model's uncertainty) \cite{Spoerer2019}. The model terminates rapidly for many images, but expends more time and energy on hard images to reach its confidence threshold. Adjusting the confidence threshold enables trading off speed for accuracy in terms of average performance. When compared to a range of FNNs requiring different amounts of computation, the RNN achieved roughly the same accuracy as each of the FNNs when its confidence threshold was set to match the FNN's computational cost (number of floating point operations) on average across images (\textbf{Fig. \ref{fig:speedacc}}). Flexible computational depth would be advantageous for animals, who may need to respond rapidly in some situations, must limit metabolic expenditures in general, and may benefit from slower and more energetically costly inferences when high accuracy is required. Computer vision faces similar requirements in certain applications. For example, a vision algorithm in a smartphone should respond rapidly and conserve energy in general, but should also be able to recognize hard images, and it should allow trading off mean accuracy for speed and energy (e.g., when the battery is low).

\subsection*{Recurrent architectures can compress complex computations in limited hardware} \noindent
Another benefit of recurrent solutions is that they require fewer components in space when physically implemented in recurrent circuits, such as brains. Compare \textbf{Figs. \ref{fig:unroll}b} and \textbf{\ref{fig:unroll}e}: the recurrent network is anatomically more compact than the feedforward network and has fewer connections. It is easy to see why evolution might have favored a recurrent implementations for many brain functions: Space, neural projections, and the energy to develop and maintain them are all costly for the organism. In addition, synaptic efficacies must be either learned from limited experience or encoded in a limited-capacity genome. Beyond saving space, material, and energy, thus, smaller descriptive complexity (or parameter count) might ease development and learning.

Engineered devices face the same set of costs, although their relative weighting changes from application to application. In particular, a larger number of units and weights must either be represented in the memory of a conventional computer or implemented in specialized (e.g., neuromorphic) hardware. The connection weights in an NN model need to be learned from limited data. This requires extensive training, e.g., in a supervised setting, with millions of hand-labeled examples that show the network the desired output for a given input. The larger number of parameters associated with a feedforward solution might overfit the training data. The learned parameters then do not generalize well to new examples of the same task. 

FNNs often turn out to generalize surprisingly well even when they have very large numbers of parameters \cite{Advani2017,Belkin2019,Nakkiran2019}. This phenomenon is thought to reflect a regularizing effect of the learning algorithm, stochastic gradient descent. Indeed, the trend is towards ever deeper networks with more connections to be optimized, and this trend is associated with continuing gains in performance on computer vision benchmarks \cite{Rawat2017}. Nevertheless, it could turn out that recurrent architectures that achieve high computational depth with fewer parameters bring benefits not only in terms of their storage, but also in terms of statistical efficiency, the ability generalize accurately based on limited experience. This would imply that recurrent networks have an inductive bias that makes up for the limited experiential data. This is explored further in subsequent sections, where we discuss how RNNs can exploit temporal dependency structures, and enable iterative inference.

Energy is another factor to consider in both biology and engineering.  Larger FNNs take longer to train on bigger computing clusters, while drawing greater amounts of power -- a trend that is not sustainable. In the long run, therefore, computer vision too may benefit from the anatomical compression that can be achieved through clever use of recurrence. 

Importantly, however, not every deep feedforward model can be compressed into an equivalent recurrent implementation. This anatomical compression can only be achieved when the same function may be applied iteratively or recursively within the network. The crucial question, therefore, is: what are these functions? What operations can be applied repeatedly in a productive manner? The remainder of this paper will reflect on the various roles that have been proposed for recurrent processing for visual inference, from superficial to increasingly profound forms of recurrence.

\subsection*{Feedback connections are required to integrate information from outside the visual hierarchy} \noindent
A key, established role of recurrent connections in biological vision is to propagate information from outside the visual cortex, so that it can aid visual inference\cite{Gilbert2013}. Here, we will briefly discuss two such outside influences: attention and expectations.

\begin{figure*}
    %\centering
    \includegraphics[width=0.65\textwidth]{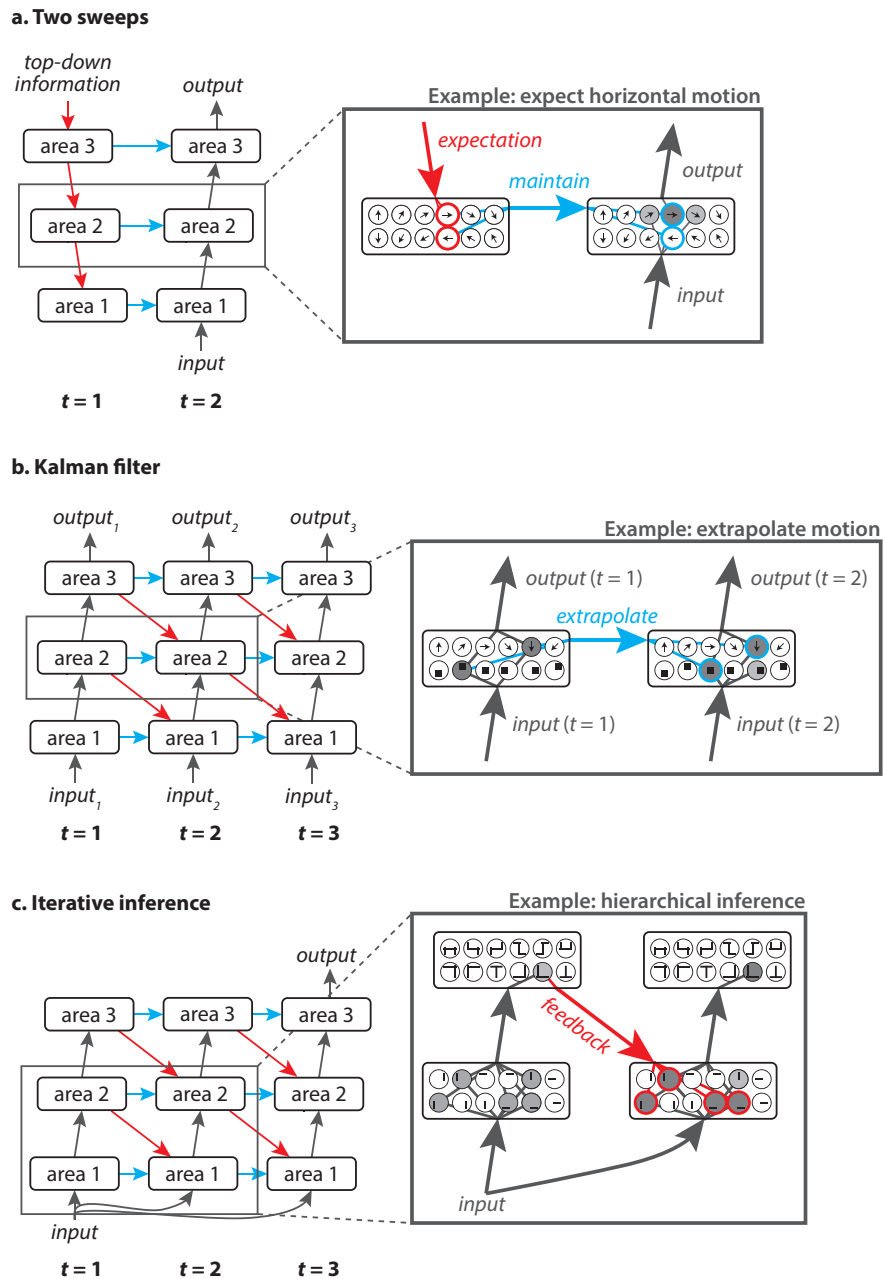}
    \caption{\textbf{Increasingly profound modes of recurrent processing, unrolled in time.} Visual cortex likely combines all three modes of recurrence illustrated here. The left side of each panel shows the computational graph induced by each form of recurrence, while the right side illustrates a (simplified) example of how this recurrence can be used. In these examples, circles correspond to neurons (or neural assemblies) encoding the feature illustrated within the circle, and lines that connect to circles indicate neural connections with significant activity. (\textbf{a}) Top-down influences from outside the visual processing hierarchy may be incorporated through two computational sweeps: a feedback sweep priming the network with top-down information and a feedforward sweep to interpret visual input and combine this interpretation with the top-down signal. Note that the lateral connections here merely copy neural activities in each area to the next time point; this identity transformation could also be implemented in other ways, such as slow membrane time constants or other forms of local memory. In the example on the right, a top-down signal communicates the expectation that the upcoming input will be horizontal motion. This primes neurons encoding this direction of motion to be more easily or strongly activated, and sharpens the interpretation of the subsequent (ambiguous) visual input. (\textbf{b}) To efficiently perform inference on time-varying visual input, recurrent connections may implement a fixed temporal prediction function akin to the transition kernel in a Kalman filter, extrapolating the ongoing dynamics of the world one time step into the future. For instance, in the example on the right, a downward moving square was perceived at $t=1$. This motion is predicted to continue, and this prediction constrains the interpretation of the (ambiguous) visual input at the next time point. For simplicity, only lateral recurrence is shown in this example. Note that each input is mapped onto its corresponding output in a single recurrent time step. (\textbf{c}) Static input may also benefit from recurrent processing that iteratively refines an initial, coarse feedforward interpretation. In this mode of recurrence, there are several processing time steps between input and output, whereas in (\textbf{b}) there was one input and output for each time step. Illustrated on the right is an iterative hierarchical inference algorithm. Here, a higher-level hypothesis, generated in the first time step, refines the underlying lower-level representation in the next time step, which in turn improves the higher-level hypothesis, and so forth, until the network converges to an optimal interpretation of the input across the entire hierarchy. For simplicity, lateral recurrent interactions are not shown in this example.}
    \label{fig:degrec}
\end{figure*}

\subsubsection*{Attentional prioritization requires feedback connections} \noindent
Animals have needs and goals that change from moment to moment. Perception is attuned to an animal's current objectives. For instance, a primate foraging for red berries may be more successful if its visual perception apparatus prioritizes or enhances the processing of red items. Since current goals are represented outside the visual cortex (e.g., in frontal regions), top-down connections are clearly required for this information to influence visual processing. Such top-down effects have been grouped under the label "attention", and they have been the subject of an entire sub-field of study. For our purposes, it is sufficient to note that the effects and mechanisms of top-down attention are well-documented and pervasive in visual cortex (for review, see [\cite{Desimone1995,Kastner2000,Maunsell2006}]), and thus there is no question that this is one important function of recurrent connections. 

\subsubsection*{Integrating prior expectations into visual inference requires feedback connections} \noindent
Organisms may constrain their visual inferences by expectations\cite{Summerfield2009}. Visual input can be ambiguous and unreliable, and thus open to multiple interpretations. To constrain the inference, an observer can make use of prior knowledge\cite{von1867handbuch,Weiss2002,Stocker2006}. One form of prior knowledge is environmental constants (e.g., "light tends to come from above"\cite{Mamassian2001}). Such unvarying knowledge may be stored within visual cortex, especially when it pertains to the overall prevalence of basic visual features (e.g., local edge orientations \cite{Girshick2011}). Another form of prior knowledge is contextual information specific to the current situation. Such time-varying knowledge may require a flexible representation outside visual cortex (e.g., "I rang the doorbell at my mother's house, so I expect to see her open the door"). Such expectations, represented in higher cortical regions, require feedback connections to affect processing in visual cortex \cite{Summerfield2009}. 

The top-down imposition of attention and expectation must be mediated by feedback connections. However, it is unclear whether these influences fundamentally change the nature of visual representations or merely modulate these representations, adjusting the gain depending on the current relevance of different features of the visual input. As illustrated in \textbf{Fig. \ref{fig:degrec}a}, for a given input this would require only two "sweeps" of computation through the visual processing hierarchy: a feedback sweep that primes visual areas with top-down information, and a bottom-up sweep to interpret the visual input and integrate or modify this interpretation with the top-down signal (not necessarily in that order). Importantly, if the feedback signal merely enhances or suppresses some visual features, then the core inference algorithm need not be fundamentally recurrent -- one can imagine that the bottom-up part of such a network is modeled perfectly by an FNN, while an optional recurrent module could be added in order to implement top-down contextual influences.

\subsection*{Recurrent networks can exploit temporal dependency structure} \noindent
Contextual constraints on visual inference include not only information from outside the visual hierarchy, such as information from other sensory modalities and memory, as discussed in the previous section. The recent stimulus history within the visual modality also provides context, likely represented within the visual system.

\subsubsection*{Recurrent networks can dynamically compress the stimulus history} \noindent
The primate visual system is thought to contain a hierarchy, not only of processing stages and spatial scales, but also of temporal scales \cite{Hasson2008,Murray2014}. Visual representations track the environment moment by moment. However, the duration of a visual moment, the temporal grain, may depend on the level of representation. These principles apply to all sensory modalities and have been empirically explored, in particular, for audition and speech perception. At the simplest level, a neural network could use delay lines to detect spatiotemporal, rather than purely spatial, patterns. Recurrent neural networks have internal states and can represent temporal context across units tuned to different latencies. An RNN could represent a fixed temporal window, by replicating units tuned to different patterns for multiple latencies. However, RNNs trained on sequence processing tasks, such as language translation, learn more sophisticated representations of temporal context \cite{Sutskever2014}. They can represent context at multiple time scales, learning a latent representation that enables them to dynamically compress whatever information from the past is needed for the task. In contrast to a feedforward network, a recurrent network is not limited by spatial constraints in terms of its retrospective time horizon. It can maintain task-relevant information indefinitely, integrating long-term memory into its inferences.

\subsubsection*{Recurrent dynamics can simulate and predict the dynamics of the world} \noindent
Dynamic compression of the past exploits the temporal dependency structure of the sensory data. The purpose of representing the past is to act well in the future. This suggests that a neural network should exploit temporal dependencies not just to compress the past, but also to predict the future. In fact, an optimal representation of even just the present requires prediction, because the sensory data is delayed and noisy.

Changes in the world are governed by laws of dynamics, which by definition are temporally invariant. An ideal observer will exploit these laws in visual inference and optimally combine previous with present observations to estimate the current state. This implies an extrapolation of the past to generate predictions that improve the interpretation of the present sensory input. When the dynamics are linear and noise is Gaussian, the optimal way to infer the present state by combining past and present evidence is the \textit{Kalman filter}\cite{Kalman1960} -- an algorithm widely used in engineering applications. A number of authors \cite{Wolpert1995,Rao1997a,Rao2004,Deneve2007} have proposed that the visual cortex may implement an algorithm similar to a Kalman filter. This theory is consistent with temporal biases that are evident in human perceptual judgments \cite{Xivry2013,Kwon2015,VanBergen2019}.

Kalman filters employ a fixed temporal transitional kernel. This kernel takes a representation of the world (e.g., variables encoding the present state of a physical system, such as positions and velocities) at time $t$, and transforms it into a predicted representation for time $t+1$, to be integrated with new sensory evidence that arrives at that time. While the resulting prediction varies as a function of the kernel's input, the kernel itself is constant, reflecting the temporal shift-invariance of the laws governing the dynamics. Recurrent neural networks provide a generalization of the Kalman filter and can represent nonlinear dynamical systems with non-Gaussian noise. 

Note that this type of recurrent processing is more profound than the two-sweep algorithm (\textbf{Fig. \ref{fig:degrec}a}) that incorporated top-down influences on visual inference. The two-sweep algorithm is trivial to unroll into a feedforward architecture. In contrast, unrolling a Kalman filter-like recurrent algorithm would induce an infinitely deep feedforward network, with a separate set of areas and connections for each time point to be processed. A finite-depth feedforward architecture can only approximate the recurrent algorithm. While the feedforward approximation will have a finite temporal window of memory to constrain its present inferences, the recurrent network can in principle integrate information over arbitrarily long periods.

Due to their advantages for dealing with time-varying (or otherwise ordered) inputs, recurrent neural networks are in fact widely employed in the broader field of machine learning for tasks involving sequential data. Speech recognition and machine translation are prominent applications that RNNs excel at\cite{Graves2013,Sak2014,Sutskever2014,Bahdanau2014,Cho2014}. Computer vision, too, has embraced RNNs for recognition and prediction of video input\cite{Ranzato2014,Srivastava2015,Lotter2016}. Note that these applications all exploit the dynamics in RNNs to model the dynamics in the data.

What if we trained a Kalman filter or sequence-to-sequence RNN (\textbf{Fig. \ref{fig:degrec}b}) on a train of independently sampled \textit{static} inputs to be classified? The memory of the preceding inputs would not be useful then, so we expect the recurrent model to revert to using essentially only its feedforward weights. The type of recurrent processing we described in this section, thus uses memory to improve visual inference. In the next section, we consider how recurrent processing can help with the inferential computations themselves, even for static inputs.

\subsection*{Recurrence enables iterative inference} \noindent
Recurrent processing can contribute even to inference on static inputs, and regardless of the agent's goals and expectations, by means of an \textit{iterative} algorithm. An iterative algorithm is one that employs a computation that improves an initial guess. Applying the computation again to the improved guess yields a further improvement. This process can be repeated until a good solution has been achieved or until we run out of time or energy. Recurrent networks can implement iterative algorithms, with the same neural network functions applied successively to some internal pattern of activity \textbf{Fig. \ref{fig:degrec}c}).

In many fields, iterative algorithms are used to solve estimation and optimization problems. In each iteration, a small adjustment is made to the problem's proposed solution, to improve a mathematically formulated objective. A locally optimal solution is found by making small improvements until further progress is not required or not possible. The algorithm navigates a path in the space of the values to be estimated or the parameters to be optimized, that leads to a good solution (albeit not necessarily the global optimum). 

Much of machine learning involves iterative methods. Gradient descent is an iterative optimization method, whose stochastic variant is the most widely used method for training FNNs. Many discrete optimization techniques are iterative. Iterative algorithms are also central to inference in machine learning, for example in variational inference (where inference is achieved by optimization), sampling methods (where steps are chosen stochastically such that the distribution of samples converges on the posterior distribution), and message passing algorithms (such as loopy belief propagation). In particular, such iterative inference algorithms are used in probabilistic approaches to computer vision \cite{Yuille2006,Prince2012}. It is somewhat surprising, then, that iterative computation is not widely exploited to perform visual inference in FNNs. 

Visual inference is naturally understood as an optimization problem, where the goal is to find hypotheses that can explain the current visual input \cite{von1867handbuch}. A hypothesis, in this case, is a proposed set of \textit{latent} (i.e. unobserved) causes that can jointly explain the image. The hypothesized latent causes could be the identities and positions of objects in the scene. Visual hypotheses are hierarchical, being subdivided into smaller hypotheses about lower or intermediate-level features, such as the local edges that make up a larger contour. An iterative visual inference algorithm starts with an initial hypothesis, and refines it by incremental improvements. These improvements may include eliminating hypotheses that are mutually exclusive, strengthening compatible causes, or adjusting a hypothesis based on its ability to predict the data (the visual input). In a probabilistic framework, the optimization objective would be the likelihood (probability of the image given the latent representation) or the posterior probability (probability of the latent representation given the image).

\subsubsection*{Incompatible hypotheses can compete in the representation} \noindent
There are often multiple plausible explanations for a given sensory input that are mutually exclusive. The distributed, parallel nature of neural networks enables them to initially activate and represent all of these possible hypotheses simultaneously. Recurrent connectivity between neurons can then implement competitive interactions among hypotheses, so as to converge on the best overall explanation.

There is some evidence that sensory representations are probabilistic \cite{Pouget2013,Ma2014,Orban2016} -- in this case, the probabilities assigned to a set of mutually exclusive hypotheses must sum to 1. A strengthening of belief in one hypothesis, thus, should entail a reduction of the probability of other hypotheses in the representation. If neurons encode point estimates rather than probability distributions, then only one hypothesis can win (although that hypothesis may be encoded by a population response involving multiple neurons). The winning hypothesis could be the maximum a posteriori (MAP) hypothesis or the maximum likelihood hypothesis. Influential models of visual inference involving competitive recurrent interactions include \textit{divisive normalization} \cite{Carandini2012}, biased competition \cite{Desimone1995}, and \textit{predictive coding} \cite{Rao1999,Friston2009,Boerlin2013}. 

Recent theoretical work has demonstrated that lateral competition can give rise to a robust neural code, and can explain certain puzzling neural response properties \cite{Boerlin2013,Barrett2016}. This theory considers a spiking neural network setting, in which different neurons encode highly overlapping or even identical features in their input. This degeneracy means that the same signal can be encoded equally well by a range of different response patterns. When a particular neuron spikes, lateral inhibition ensures that other competing neurons do not encode the same part of the input again. Which neuron gets to do the encoding thus depends on which neuron fires first, because its membrane potential happened to be closest to a spiking threshold. This leads to trial-to-trial variability in neural responses that reflects subtle differences in initial conditions -- conditions that may not be known to an experimenter, who may thus mistake this variability for random noise. This could explain the puzzling observation that individual neurons reliably reproduce the same output given the same electrical stimulation, but populations of neurons, wired together, display apparently random variability under sensory stimulation \cite{Schiller1976,Dean1981,Mainen1995}. Since multiple neurons can encode the same feature, the resulting code is also robust to neurons being lost or temporarily inactivated. 

FNNs do not incorporate lateral connections for competitive interactions, although they very often include computations that serve a similar purpose. Chief among these are operations known as \textit{max-pooling} and \textit{local response normalization} (LRN) \cite{Krizhevsky2012,Lecun2015}. In max-pooling, only the strongest response within a pool of competing neurons is forwarded to the next processing stage. In LRN, each neuron has its response divided by a term that is computed from the sum of activity in its normalization pool. While neither of these mechanisms is mediated by explicit lateral connections in a FNN, a strictly connectionist implementation of these mechanisms (e.g., in biological neurons or neuromorphic hardware) would have to include lateral recurrence. This, then, is another way in which apparently feedforward FNNs can exhibit a (limited) form of recurrent processing "under the hood". Note, though, that each of these operations is carried out only once, rather than allowing competitive dynamics to converge over multiple iterations. Furthermore, in contrast to the lateral interactions in predictive coding or other normative models, LRN and max-pooling are not derived from normative principles, and do not necessarily select (or enhance) the best hypothesis (however "best" is defined).

\subsubsection*{Compatible hypotheses can strengthen each other in the representation} \noindent
In feedforward models of hierarchical visual inference, neurons at higher stages selectively respond to combinations of simpler features encoded by lower-level neurons. Higher-level neurons thus are sensitive to larger-scale patterns of correlation between subsets of lower-level features. But such larger-scale statistical regularities may not be most efficiently captured by a set of larger-scale building blocks. Instead, they may be more compactly captured by local association rules. Consider, for instance, the problem of contour detection. Many combinations of local edges in an image can form a continuous contour. The resulting space of contours may be too complex to be efficiently represented with larger-scale templates. What all these contours have in common, however, is that they consist of pairs of edges that are locally contiguous, with sharper angles occurring with lower probability. Thus, the criteria for 'contour-ness' may be compactly expressed by a set of local association rules: \textit{these edges go together; those do not}\cite{Field1993,Geisler2001}. Contours may then be pieced together by repeatedly applying the same local association rules. Those edge pairs which are most clearly connected would be identified in early iterations. Later inferences can benefit from the context provided by earlier inferences, enabling the process to recognize continuity even where it is less locally apparent.

This insight has inspired network models of visual inference that implement local association rules through lateral connections, to aid contour integration and other perceptual grouping operations\cite{Roelfsema2006}. Recent examples include Linsley et al., who developed \textit{horizontal gated-recurrent units} (hGRUs) that learn local spatial dependencies \cite{Linsley2018b}. A network equipped with this particular recurrent connectivity was competitive with state-of-the-art feedforward models on a contour integration task, while using far fewer free parameters. George et al. \cite{George2017} similarly leveraged lateral interactions to recognize contiguous contours and surfaces, by modeling these with a conditional random field (CRF), using a message-passing algorithm for inference. This approach made their Recursive Cortical Network (RCN) the first computer vision algorithm to reliably beat CAPTCHAs -- images of letter sequences under a variety of distortions, noise and clutter, that are widely used to verify that queries to a user interface are made by a person, and not an algorithm. CRFs were also used by Zheng et al. \cite{Zheng2015}, who incorporated them as a recurrent extension of a convolutional neural network for image segmentation. The model surpassed state-of-the-art performance at the time. Association rules enforced through lateral connections may also help to fill in missing information, such as when objects are partially hidden from view by occluders. Lateral connectivity has been shown to improve recognition performance in such settings \cite{Spoerer2017,Spoerer2019,Montobbio2019a}. Montobbio et al. showed that lateral diffusion of activity between neurons with correlated feedforward filter weights improves robustness to image perturbations including occlusions\cite{Montobbio2019a}. 

Enhancement of mutually compatible hypotheses (this section) and competition between mutually exclusive hypotheses (previous section) can both contribute to inference. A more general perspective is provided by the insight that prior knowledge about what features in a scene are mutually compatible or exclusive may be part of an overarching \textit{generative model}, which iterative algorithms can exploit for inference. 

\subsubsection*{Iterative algorithms can leverage generative models for inference} \noindent
Perceptual inference aims to converge on a set of hypotheses that best explain the sensory data. Typically, a hypothesis is considered to be a good explanation if it is consistent with both our prior knowledge and the sensory data. A generative model is a model of the joint distribution of latent causes and sensory data. Generative models can powerfully constrain perceptual inference because they capture prior knowledge about the world. In machine learning, defining generative models enables us to express and exploit what we know about the domain. A wide range of inference algorithms can be used to compute posterior distributions over variables of interest, given observed variables. The algorithms include variational inference, message passing, and Markov Chain Monte Carlo sampling, all of which require iterative computation. 

In this section, we focus on a particular approach to leveraging generative models in visual inference, in which the joint distribution $p(x,z)$ of the image $x$ and the latents $z$ is factorized as $p(x,z) = p(z)\cdot p(x|z)$, which we refer to as the top-down factorization. The architecture contains components that model $p(x|z)$ and predict the image from the latents (or more generally lower-level latent representations from higher-level latent representations). Compared to the alternative factorization $p(x,z) = p(x)\cdot p(z|x)$, the top-down factorization has the potential advantage that the model operates in the causal direction, matching the causal process in the world that generated the image. The top-down model predicts what visual input is likely to result from a scene that has the hypothesized properties. This is somewhat similar to the graphics engine of a video game or image rendering software. This top-down model can be implemented via feedback connections that translate higher-level hypotheses in the network to representations at a lower level of abstraction.

Using generative models implemented with top-down predictions for inference is known as \textit{analysis-by-synthesis} -- an approach that has a long history in theories of perception \cite{von1867handbuch,Rao1999,Friston2009}. Arguably, the goal of perceptual inference, by definition, is to reason back from effects (sensory data) to their causes (unobserved variables of interest), and thus invert the process that generated the effects. The crucial question, however, is whether the causal process is explicitly represented in the inference algorithm. The alternative, which can be achieved with feedforward inference, is to directly approximate the inverse, without ever making predictions in the causal direction. The success of the feedforward approach then depends on how well the inverse can be approximated by a fixed mapping of inputs to hypotheses. To iteratively invert the causal process, a neural network can evaluate the causal model for a current hypothesis and update the hypothesis in a beneficial direction. This process can then be repeated until convergence. This process of analysis by repeated synthesis may be preferable to directly approximating the inverse mapping if the causal process that generates the sensory data is easier to model than its inverse. In particular, the causal process may be more compactly represented, more easily learned, more efficient to compute, and more generalizable beyond the training distribution than its inverse. 

Another potential advantage of generative inference lies in robustness to variations in the input. While FNNs can accurately categorize images drawn from the same distribution that the training images were drawn from, it does not take much to fool them. A slight alteration imperceptible to humans can cause a FNN to misclassify an image entirely, with high confidence \cite{Szegedy2014}. State-of-the-art FNNs rely more strongly on texture than humans, who rely more on shape \cite{Geirhos2019}. More generally, FNNs seem to ignore many image features that are relevant to human perception \cite{Jacobsen2019}. One hypothesized reason for this is that these networks are trained to discriminate images, but not to generate them. Thus, any visual feature that reliably discriminates categories in the training data will be weighted heavily in the network's classification decisions. Importantly, this weight is unrelated to how much variance the feature explains in the image, and to the likelihood, i.e. the probability of the image given either of the categories. An ideal observer should evaluate the likelihood for each hypothesis and adjudicate according to their ratio \cite{Neyman1933}. A feedforward network may instead latch on to a few highly discriminative, but subtle image features that don't explain much and may not generalize to images from a different data set \cite{Jacobsen2019,Engstrom2019}. In contrast, visual features that are important for generating or reconstructing images of a given class may be more likely to generalize to other examples of the same category. In support of this intuition, two novel RNN architectures that employ generative models for inference were found to be more robust to adversarial perturbations \cite{Li2019,Schott2018}. Generative inference networks were also shown to better align with human perception, compared to discriminative models, when presented with \textit{controversial stimuli} -- images synthesized to evoke strongly conflicting classifications from different models \cite{Golan2019}. 

Despite these promising developments, generative inference remains rare in visual FNN models. The exceptions mentioned above are rather simple networks trained on easy classifications problems, and are not (yet) competitive with state-of-the-art performance on more challenging computer vision benchmarks. Within computational neuroscience, by contrast, generative feedback connections appear in many network models of visual inference. Prominent examples are predictive coding \cite{Rao1999,Friston2009} and hierarchical Bayesian inference \cite{Lee2003}. However, these models have not had much success in explaining visual inference beyond its earliest stages. A notable exception is work by Wen et al. \cite{Wen2018}, which shows that extending supervised convolutional FNNs with the recurrent dynamics of predictive coding can improve classification performance. The fields of computer vision and computational neuroscience both stand to benefit from the development of more powerful generative inference models.

\subsubsection*{Iteration is necessary to close the amortization gap} \noindent
Iterative inference has many advantages. A drawback of iteration, however, is that it takes time for the algorithm to converge during inference. This is unattractive for animals who need to perform visual inference under time pressure. It is also a challenge when training a FNN, which already requires many iterations of optimization. If each update of the network's connections additionally includes an iterative inner loop to perform inference on each training example, this lengthens the time required for training.

A complementary inference mechanism is \textit{amortized inference}\cite{Srikumar2012,Stuhlmuller2013}, where a feedforward model approximates the mapping from images to their latent causes. FNNs are eminently suited for learning complicated input-output mappings. A single transformation then replaces the trajectories that would be navigated by an iterative inference algorithm. In some cases, the iterative solution and the best amortized mapping may be exactly equivalent. A linear model, for instance, can be estimated iteratively, by performing gradient descent on the sum of squared prediction errors. However, if a unique solution exists, it can equivalently be found by a linear transformation that directly maps from the data to the optimal coefficients.

In general, however, amortized inference incurs some error, compared to the optimal solution that might be found through iterative optimization. This error has been called the \textit{amortization gap} \cite{Cremer2018,Marino2018}. It is analogous to the poor fit that may result from buying clothes "off the rack", compared to a tailored version of the same garment. The amortization gap is defined in the context of variational inference, when the iterative optimization of the variational approximation to the posterior is replaced by a neural network that maps from the image to the parameters of the variational distribution. The resulting model suffers from two types of error: (1) error caused be the choice of the variational approximation (variational approximation gap) and (2) error caused by the model mapping from images to variational parameters (amortization gap). One recent study has argued that the amortization gap is often the main source of error in amortized inference models \cite{Cremer2018}.
 
Amortized and iterative inference define a continuum. At one extreme, iterative inference until convergence reaches a solution through a trajectory of small improvements, explicitly evaluating the quality of the current solution at every iteration. At the other extreme, fully amortized inference takes a single leap from input to output. In between these extremes lies a space for algorithms that use intermediate numbers of steps, to approximate the optimal solution through a computational path that is more refined than a leap, but more efficient than full-fledged iterative optimization. Models that occupy this space include explicit hybrids of iterative and amortized inference \cite{Hjelm2016,Krishnan2018,Marino2018}, as well as RNNs with arbitrary dynamics that are trained to converge to a desired objective in a limited number of time steps (e.g., \cite{Liang2015,Spoerer2019,Kar2019,Nayebi2018}). 

\subsection*{Recurrence is required for active vision} \noindent
Vision is an active exploratory process. Our eye movements scan the scene through a sequence of well-chosen fixations that bring objects of interest into foveal vision. Moving our heads and our bodies enables us to bring entirely new parts of the scene into view, and closer for inspection at high resolution. Active control of our eyes, heads, and bodies can also help disambiguate 3D structure as fixation on points at different depths changes binocular disparity, and head and body movements create motion parallax. Active vision involves a recurrent cycle of sensory processing and muscle control, a cycle that runs through the environment. 

Our focus here has been on the internal computational functions of recurrent processing, and active vision has been reviewed elsewhere \cite{Ballard1991,Findlay2003,Bajcsy2018}. However, it is important to note that the internal recurrent processes of visual inference from a single glimpse are embedded within the larger recurrent process of active visual exploration. Active vision provides not just the larger behavioral context of visual inference. It also provides a powerful illustration of the fundamental advantages that recurrent algorithms offer in general. It illustrates how limited resources (the fovea) can be dynamically allocated (eye movements) to different portions of the evidence (the visual scene) in temporal sequence. A sensory system limited to a finite number of neurons, thus, can multiply its resources along time to achieve a detailed analysis. The cycle may start with an initial rough analysis of the entire visual field, followed by fixations on locations likely to yield valuable information. This is an example of an essentially recurrent process whose efficiency cannot be emulated with a feedforward system. The internal mechanisms of visual inference are faced with qualitatively similar challenges: Just like our retinae cannot afford foveal resolution throughout the visual field, the ventral stream cannot afford to perform all potentially relevant inferences on the evidence streaming in through the optic nerve in a single feedforward sweep. Internal shifts of attention, like eye movements, can sequentialize a complex computation and avoid wasting energy on portions of the evidence that are uninformative or irrelevant to the current goals of the animal. 

Whereas the outer loop of active vision is largely about positioning our eyes relative to the scene and bringing important content into foveal vision, the inner loop of visual inference on each glimpse is far more flexible. Beyond covert attentional shifts that select locations, features, or objects for scrutiny, a recurrent network can \textit{decide what computations to perform} so as to most efficiently reduce uncertainty about the important parts of the scene. In a game of twenty questions, we choose a question that most reduces our remaining uncertainty at each step. The budget of twenty would not suffice if we had to decide all the questions before seeing any answers. The visual system similarly has limited computational resources for processing a massive stream of evidence. It must choose what inferences to pursue on the basis of their computational cost and uncertainty-reducing benefit as it forages for insight \cite{Russell1997,Gershman2015,Griffiths2015}.

\section*{Closing the gap between biological and artificial vision} \noindent
We have reviewed a number of advantages that recurrence can bring to neural networks for visual inference. Going forward, neural network models of vision should incorporate recurrence; not just to better understand visual inference in the brain, but also to improve its implementation in machines.

\subsection*{Recurrence already improves performance on challenging visual tasks} \noindent
Efforts in this direction are already underway, and turning up promising results. Some of this work has been described in previous sections, such as the use of lateral connections to impose local association rules \cite{Linsley2018b,George2017,Zheng2015} and generative inference for more robust performance outside the training distribution\cite{Li2019,Schott2018}. Several other recent findings are worth highlighting here, as they have shown improved performance on visual tasks, better approximations to biological vision, or both, through recurrent computations. 

In particular, several studies have found that recurrence is required in order to explain or improve visual inference in challenging settings. Kar and colleagues \cite{Kar2019} identified a set of 'challenge images' that required recurrent processing in order to be accurately recognized. A feedforward FNN struggled to interpret these images, whereas macaque monkeys recognized them as accurately as a set of control images. Challenge images were associated with longer processing times in the macaque inferior temporal (IT) cortex, consistent with recurrent computations. Neural responses in IT for images that took longer were well accounted for by a brain-inspired RNN model. In a different study\cite{Kubilius2019}, this same recurrent architecture was found to account for behavior, and neural data from macaque visual cortex, in object recognition tasks, while also achieving good performance on an important computer vision benchmark (ImageNet\cite{Deng2009}). In human visual cortex, recurrent interactions were also found to be crucial to model the neural dynamics underlying object recognition, as measured through magnetoencephalography (MEG)\cite{Kietzmann2019}. 

One prominent challenge to visual inference is posed by partial occlusions, which hide part of a target object from view. In two recent studies, recurrent architectures were shown to be more robust to occlusions than their feedforward counterparts \cite{Spoerer2017,Tang2018}. Interestingly, in both human observers and in an RNN model, object recognition under occlusion was impaired by \textit{backward masking} \cite{Tang2018} (the presentation of a meaningless noise image, shortly after a target stimulus, to disrupt recurrent processing\cite{Enns2000,Fahrenfort2007,Lamme2001}). Neural responses to partially occluded shapes in macaque visual cortex are also consistent with recurrent processing, and were well explained by a predictive coding model in which prefrontal cortex provide a feedback signal to visual area V4 \cite{Fyall2017,Choi2018}. 

Another challenge for human perception is \textit{crowding}, which occurs when the detailed perception of a target stimulus is disrupted by nearby flanker stimuli\cite{Levi2008}. In certain instances, the target stimulus can be released from crowding if further flankers are added that form a larger, coherent structure with the original flankers. This \textit{uncrowding} effect may be due to the flankers being 'explained away', thus reducing their interference with the target representation\cite{Manassi2012,Manassi2016}. Recent work\cite{Doerig2020} has shown that both effects can be explained by architectures known as \textit{Capsule Nets}\cite{Sabour2017,Sabour2018}, which include recurrent information routing mechanisms that may be similar to perceptual grouping and segmentation processes in the visual cortex.

Note that, in all of these cases, it may be possible to develop a feedforward architecture that performs the task equally well or better. Trivially, and as we discussed previously, a successful recurrent architecture can always be unrolled (for a finite number of time steps) into a deep feedforward network with many more learnable connections. However, a \textit{realistic} recurrent model, when unrolled, may map onto an \textit{unrealistic} feedforward model (\textbf{Fig. \ref{fig:setmap}}), where realism could refer to the real-world constraints faced by either biological or artificial visual systems. Future studies should compare RNN and FNN implementations for the same visual inference task, while matching the complexity of the models in a meaningful way. Setting a realistic budget of units, connections, and computational operations is one important approach. To understand the computational differences between RNN and FNN solutions, it is also interesting to (1) match the parameter count (number of connection weights that must be learned and stored), which requires granting the FNN larger feature kernels, more feature maps per layer, or more layers, or (2) match the computational graph, which equates the distribution of path lengths from input to output and all other statistics of the graph, but grants the FNN a much larger number of parameters \cite{Spoerer2019}.

\subsection*{Freeing ourselves from the feedforward framework} \noindent
Deep feedforward neural networks constitute an essential building block for visual inference, but they are not the whole story. The missing element, recurrent dynamics, is central to a range of alternative conceptions of visual inference that have been proposed \cite{Ballard1991,ORegan2001,Yuille2006,Findlay2003,Bajcsy2018,Buzsaki2019}. These ideas have a long history, they are essential to understanding biological vision, and they have great potential for engineering, especially in the context of modern hardware and software. The promise of active vision and recurrent visual inference is, in fact, boosted by the power of feedforward networks. 

However, the beauty, power, and simplicity of feedforward neural networks also makes it difficult to engage and develop the space of recurrent neural network algorithms for vision. The feedforward framework, embellished by recurrent processes that serve auxiliary and modulatory functions like normalization and attention, enables computational neuroscientists to hold on to the idea of a hierarchy of feature detectors. This idea might not be entirely mistaken. However, it is likely to be severely incomplete and ultimately limiting.

The insight that any finite-time recurrent network can be unrolled compounds the problem by suggesting that the feedforward framework is essentially complete. More practically, the fact that we train RNNs by unrolling them for finite time steps might in some ways impede our progress. FNNs are usually trained by stochastic gradient descent using the \textit{backpropagation} algorithm. This method retraces in reverse the computational steps that led to the response in the output layer, so as to estimate the influence that each connection in the network had on the response. Each connection weight is then adjusted, to bring the network output closer to a desired output. The deeper the network, the longer the computational path that needs to be retraced. RNNs for visual inference typically are trained through a variation on this method, known as \textit{backpropagation through time} (BPTT)\cite{Werbos1990}. To retrace computations in reverse through cycles, the RNN is unrolled along time, so as to convert it into a feedforward network whose depth depends on the number of time steps as shown in \textbf{Fig. \ref{fig:unroll}b-d}. This enables the RNN to be trained like an FNN.

BPTT is attractive for enabling us to train RNNs like FNNs on arbitrary objectives. When it comes to learning recurrent dynamics, however, BPTT strictly optimizes the output at the specific time points evaluated by the objective (e.g., the output after exactly $N$ steps). Outside of this time window, there is no guarantee that the network's response will be well-behaved. The RNN might reach the desired objective at the desired time, but diverge immediately after. Ideally, we would like a visual RNN presented with a stable image to converge to an attractor that represents the image and behave stably for arbitrary lengths of time. This would be consistent with iterative optimization, in which each step improves the network's approximation to its objective. While it is not impossible for BPTT to give rise to such dynamics, it does not specifically favor them.

From a theory perspective, BPTT is limiting because it shackles RNNs to the feedforward framework, in which the goal is still to map inputs to outputs, rather than to discover useful dynamics. From a practical and implementational perspective, BPTT is computationally cumbersome, as every additional recurrent time step extends the computational path that must be retraced in order to update the connections. This complication also renders BPTT biologically implausible. Although the case for backpropagation as potentially biologically plausible has recently been strengthened 
\cite{Guerguiev2017,NIPS2018_8089,Whittington2019}, its extension \textit{through time} is difficult to reconcile with biology\cite{Lillicrap2019} or implement efficiently in a finite engineered system for online learning -- precisely because it requires unrolling and keeping track of separate copies of each weight as computational cycles are retraced in reverse.

Given these drawbacks, we speculate that a true breakthrough in recurrent vision models will require a training regime that does not rely on BPTT. Rather than optimizing an RNN's state in a finite time window, future RNN training methods might directly target the network's dynamics, or the states that those dynamics are encouraged to converge to. This approach has some history in RNN models of vision. Predictive coding models, for instance, are designed with dynamics that explicitly implement iterative optimization. Such models can update their connections through learning rules that require only the converged network state as input\cite{Rao1999}, rather than the entire computational path to this state. Marino et al. \cite{Marino2018} recently proposed \textit{iterative amortized inference}, training inference networks to have recurrent dynamics that improve the network's hypotheses in each iteration, without constraining these dynamics to a particular form (such as predictive coding). More generally, RNNs whose dynamics converge to a steady state can be optimized through variations on an algorithm known as \textit{recurrent backpropagation} \cite{Almeida1987,Pineda1987,Liao2018}, which avoids retracing the computational graph through time. However, it is often difficult to design RNNs such that their dynamics converge to a steady state (within the time window for which the model is trained), while maintaining \textit{expressivity} (the ability of the model to learn a wide range of functions). This challenge is addressed by the recently developed \textit{contractor recurrent backpropagation} method\cite{Linsley2020}, which introduces a mathematical penalty that can be imposed while training any RNN, to encourage it to learn convergent dynamics.
\\
\section*{Going forward, in circles} \noindent
We started this review with the puzzling observation that, whereas biological vision is implemented in a profoundly recurrent neural architecture, the most successful neural network models of vision to date are feedforward. We have argued, theoretically and empirically, that vision models will eventually converge to their biological roots and implement more powerful recurrent solutions. This is an appealing prospect, as it suggests that neuroscientists and engineers can continue to work synergistically, to make progress on common challenges. After all, visual inference, and intelligence more generally, were solved once before, and so discovering nature's solutions should go hand in hand with building artificial ones.

\section*{Acknowledgements} \noindent
We thank Samuel Lippl, Heiko Schütt, Andrew Zaharia, Tal Golan and Benjamin Peters for detailed comments on a draft of this paper. This work was supported by a Rubicon grant from the Dutch Research Council (to R.S.v.B.).

\bibliography{bib}

\begin{thebibliography}{100}
\expandafter\ifx\csname url\endcsname\relax
  \def\url#1{\texttt{#1}}\fi
\expandafter\ifx\csname urlprefix\endcsname\relax\def\urlprefix{}\fi
\expandafter\ifx\csname href\endcsname\relax
  \def\href#1#2{#2} \def\path#1{#1}\fi

\bibitem{Lamme2000}
V.~A. Lamme, P.~R. Roelfsema, {The distinct modes of vision offered by
  feedforward and recurrent processing}, Trends in Neurosciences 23~(11) (2000)
  571--579.
\newblock \href {https://doi.org/10.1016/S0166-2236(00)01657-X}
  {\path{doi:10.1016/S0166-2236(00)01657-X}}.

\bibitem{Kreiman}
G.~Kreiman, T.~Serre, {Beyond the feedforward sweep: feedback computations in
  the visual cortex}, Annals of the New York Academy of Sciences 1464~(1)
  (2020) 222--241.
\newblock \href {https://doi.org/10.1111/nyas.14320}
  {\path{doi:10.1111/nyas.14320}}.

\bibitem{Angelucci2006}
A.~Angelucci, P.~C. Bressloff, {Contribution of feedforward, lateral and
  feedback connections to the classical receptive field center and
  extra-classical receptive field surround of primate V1 neurons}, in: Progress
  in Brain Research, Vol. 154, 2006, pp. 93--120.
\newblock \href {https://doi.org/10.1016/S0079-6123(06)54005-1}
  {\path{doi:10.1016/S0079-6123(06)54005-1}}.

\bibitem{Anderson1994}
J.~C. Anderson, R.~J. Douglas, K.~A.~C. Martin, J.~C. Nelson, {Synaptic output
  of physiologically identified spiny stellate neurons in cat visual cortex},
  The Journal of Comparative Neurology 341~(1) (1994) 16--24.
\newblock \href {https://doi.org/10.1002/cne.903410103}
  {\path{doi:10.1002/cne.903410103}}.

\bibitem{Martin2002}
K.~A. Martin, {Microcircuits in visual cortex}, Current Opinion in Neurobiology
  12~(4) (2002) 418--425.
\newblock \href {https://doi.org/10.1016/S0959-4388(02)00343-4}
  {\path{doi:10.1016/S0959-4388(02)00343-4}}.

\bibitem{Douglas2007}
R.~J. Douglas, K.~A. Martin, {Recurrent neuronal circuits in the neocortex},
  Current Biology 17~(13) (2007) 496--500.
\newblock \href {https://doi.org/10.1016/j.cub.2007.04.024}
  {\path{doi:10.1016/j.cub.2007.04.024}}.

\bibitem{Felleman1991}
D.~J. Felleman, D.~C. {Van Essen}, {Distributed hierarchical processing in the
  primate cerebral cortex}, Cerebral Cortex 1~(1) (1991) 1--47.
\newblock \href {https://doi.org/10.1093/cercor/1.1.1}
  {\path{doi:10.1093/cercor/1.1.1}}.

\bibitem{Salin1995}
P.~A. Salin, J.~Bullier, {Corticocortical connections in the visual system:
  structure and function}, Physiological Reviews 75~(1) (1995) 107--154.
\newblock \href {https://doi.org/10.1152/physrev.1995.75.1.107}
  {\path{doi:10.1152/physrev.1995.75.1.107}}.

\bibitem{Markov2014}
N.~T. Markov, M.~M. Ercsey-Ravasz, A.~R. {Ribeiro Gomes}, C.~Lamy, L.~Magrou,
  J.~Vezoli, P.~Misery, A.~Falchier, R.~Quilodran, M.~A. Gariel, J.~Sallet,
  R.~Gamanut, C.~Huissoud, S.~Clavagnier, P.~Giroud, D.~Sappey-Marinier,
  P.~Barone, C.~Dehay, Z.~Toroczkai, K.~Knoblauch, D.~C. {Van Essen},
  H.~Kennedy, {A weighted and directed interareal connectivity matrix for
  macaque cerebral cortex}, Cerebral Cortex 24~(1) (2014) 17--36.
\newblock \href {https://doi.org/10.1093/cercor/bhs270}
  {\path{doi:10.1093/cercor/bhs270}}.

\bibitem{Douglas1995}
R.~J. Douglas, C.~Koch, M.~Mahowald, K.~A. Martin, H.~H. Suarez, {Recurrent
  excitation in neocortical circuits}, Science 269~(5226) (1995) 981--985.
\newblock \href {https://doi.org/10.1126/science.7638624}
  {\path{doi:10.1126/science.7638624}}.

\bibitem{Super2001}
H.~Sup{\`{e}}r, H.~Spekreijse, V.~A. Lamme, {Two distinct modes of sensory
  processing observed in monkey primary visual cortex (VI)}, Nature
  Neuroscience 4~(3) (2001) 304--310.
\newblock \href {https://doi.org/10.1038/85170} {\path{doi:10.1038/85170}}.

\bibitem{DiLollo2000}
V.~{Di Lollo}, J.~T. Enns, R.~A. Rensink, {Competition for consciousness among
  visual events: The psychophysics of reentrant visual processes}, Journal of
  Experimental Psychology: General 129~(4) (2000) 481--507.
\newblock \href {https://doi.org/10.1037/0096-3445.129.4.481}
  {\path{doi:10.1037/0096-3445.129.4.481}}.

\bibitem{Lamme2001}
V.~A. Lamme, K.~Zipser, H.~Spekreijse, {Masking interrupts figure-ground
  signals in V1}, Journal of Vision 1~(3) (2001) 1044--1053.
\newblock \href {https://doi.org/10.1167/1.3.32} {\path{doi:10.1167/1.3.32}}.

\bibitem{Heinen2005}
K.~Heinen, J.~Jolij, V.~A. Lamme, {Figure-ground segregation requires two
  distinct periods of activity in VI: A transcranial magnetic stimulation
  study}, NeuroReport 16~(13) (2005) 1483--1487.
\newblock \href {https://doi.org/10.1097/01.wnr.0000175611.26485.c8}
  {\path{doi:10.1097/01.wnr.0000175611.26485.c8}}.

\bibitem{Fahrenfort2007}
J.~J. Fahrenfort, H.~S. Scholte, V.~A. Lamme, {Masking disrupts reentrant
  processing in human visual cortex}, Journal of Cognitive Neuroscience 19~(9)
  (2007) 1488--1497.
\newblock \href {https://doi.org/10.1162/jocn.2007.19.9.1488}
  {\path{doi:10.1162/jocn.2007.19.9.1488}}.

\bibitem{Lecun2015}
Y.~Lecun, Y.~Bengio, G.~Hinton, {Deep learning}, Nature 521~(7553) (2015)
  436--444.
\newblock \href {https://doi.org/10.1038/nature14539}
  {\path{doi:10.1038/nature14539}}.

\bibitem{Schmidhuber2015}
J.~Schmidhuber, {Deep learning in neural networks: An overview}, Neural
  Networks 61 (2015) 85--117.
\newblock \href {https://doi.org/10.1016/j.neunet.2014.09.003}
  {\path{doi:10.1016/j.neunet.2014.09.003}}.

\bibitem{He2015}
K.~He, X.~Zhang, S.~Ren, J.~Sun, {Delving Deep into Rectifiers: Surpassing
  Human-Level Performance on ImageNet Classification}, in: 2015 IEEE
  International Conference on Computer Vision (ICCV), Vol. 2015 Inter, IEEE,
  2015, pp. 1026--1034.
\newblock \href {https://doi.org/10.1109/ICCV.2015.123}
  {\path{doi:10.1109/ICCV.2015.123}}.

\bibitem{He2016}
K.~He, X.~Zhang, S.~Ren, J.~Sun, {Deep residual learning for image
  recognition}, Proceedings of the IEEE Computer Society Conference on Computer
  Vision and Pattern Recognition 2016-December (2016) 770--778.
\newblock \href {https://doi.org/10.1109/CVPR.2016.90}
  {\path{doi:10.1109/CVPR.2016.90}}.

\bibitem{Kemelmacher-Shlizerman2016}
I.~Kemelmacher-Shlizerman, S.~M. Seitz, D.~Miller, E.~Brossard, {The MegaFace
  benchmark: 1 million faces for recognition at scale}, Proceedings of the IEEE
  Computer Society Conference on Computer Vision and Pattern Recognition
  2016-December (2016) 4873--4882.
\newblock \href {http://arxiv.org/abs/1512.00596} {\path{arXiv:1512.00596}},
  \href {https://doi.org/10.1109/CVPR.2016.527}
  {\path{doi:10.1109/CVPR.2016.527}}.

\bibitem{Kubilius2016}
J.~Kubilius, S.~Bracci, H.~P. {Op de Beeck}, {Deep Neural Networks as a
  Computational Model for Human Shape Sensitivity}, PLOS Computational Biology
  12~(4) (2016) e1004896.
\newblock \href {https://doi.org/10.1371/journal.pcbi.1004896}
  {\path{doi:10.1371/journal.pcbi.1004896}}.

\bibitem{Majaj2018}
N.~J. Majaj, D.~G. Pelli, {Deep learning-Using machine learning to study
  biological vision}, Journal of Vision 18~(13) (2018) 1--13.
\newblock \href {https://doi.org/10.1167/18.13.2} {\path{doi:10.1167/18.13.2}}.

\bibitem{Spoerer2019}
C.~J. Spoerer, T.~C. Kietzmann, J.~Mehrer, I.~Charest, N.~Kriegeskorte,
  {Recurrent neural networks can explain flexible trading of speed and accuracy
  in biological vision}, PLOS Computational Biology 16~(10) (2020) e1008215.
\newblock \href {https://doi.org/10.1371/journal.pcbi.1008215}
  {\path{doi:10.1371/journal.pcbi.1008215}}.

\bibitem{Cadieu2014}
C.~F. Cadieu, H.~Hong, D.~L.~K. Yamins, N.~Pinto, D.~Ardila, E.~A. Solomon,
  N.~J. Majaj, J.~J. DiCarlo, {Deep Neural Networks Rival the Representation of
  Primate IT Cortex for Core Visual Object Recognition}, PLoS Computational
  Biology 10~(12) (2014) e1003963.
\newblock \href {https://doi.org/10.1371/journal.pcbi.1003963}
  {\path{doi:10.1371/journal.pcbi.1003963}}.

\bibitem{Khaligh-Razavi2014}
S.~M. Khaligh-Razavi, N.~Kriegeskorte, {Deep Supervised, but Not Unsupervised,
  Models May Explain IT Cortical Representation}, PLoS Computational Biology
  10~(11) (2014).
\newblock \href {https://doi.org/10.1371/journal.pcbi.1003915}
  {\path{doi:10.1371/journal.pcbi.1003915}}.

\bibitem{Guclu2015}
U.~Guclu, M.~A.~J. van Gerven, {Deep Neural Networks Reveal a Gradient in the
  Complexity of Neural Representations across the Ventral Stream}, Journal of
  Neuroscience 35~(27) (2015) 10005--10014.
\newblock \href {https://doi.org/10.1523/JNEUROSCI.5023-14.2015}
  {\path{doi:10.1523/JNEUROSCI.5023-14.2015}}.

\bibitem{Kriegeskorte2015}
N.~Kriegeskorte, {Deep Neural Networks: A New Framework for Modeling Biological
  Vision and Brain Information Processing}, Annual Review of Vision Science
  1~(1) (2015) 417--446.
\newblock \href {https://doi.org/10.1146/annurev-vision-082114-035447}
  {\path{doi:10.1146/annurev-vision-082114-035447}}.

\bibitem{Kheradpisheh2016}
S.~R. Kheradpisheh, M.~Ghodrati, M.~Ganjtabesh, T.~Masquelier, {Deep Networks
  Can Resemble Human Feed-forward Vision in Invariant Object Recognition},
  Scientific Reports 6~(1) (2016) 32672.
\newblock \href {https://doi.org/10.1038/srep32672}
  {\path{doi:10.1038/srep32672}}.

\bibitem{Schrimpf2018}
M.~Schrimpf, J.~Kubilius, H.~Hong, N.~J. Majaj, R.~Rajalingham, E.~B. Issa,
  K.~Kar, P.~Bashivan, J.~Prescott-Roy, F.~Geiger, K.~Schmidt, D.~L.~K. Yamins,
  J.~J. DiCarlo, Brain-score: Which artificial neural network for object
  recognition is most brain-like?, bioRxiv (2020).
\newblock \href {https://doi.org/10.1101/407007} {\path{doi:10.1101/407007}}.

\bibitem{Rao1999}
R.~P.~N. Rao, D.~H. Ballard, {Predictive coding in the visual cortex: a
  functional interpretation of some extra-classical receptive-field effects.},
  Nature neuroscience 2~(1) (1999) 79--87.
\newblock \href {https://doi.org/10.1038/4580} {\path{doi:10.1038/4580}}.

\bibitem{Yuille2006}
A.~Yuille, D.~Kersten, {Vision as Bayesian inference: analysis by synthesis?},
  Trends in Cognitive Sciences 10~(7) (2006) 301--308.
\newblock \href {https://doi.org/10.1016/j.tics.2006.05.002}
  {\path{doi:10.1016/j.tics.2006.05.002}}.

\bibitem{Friston2009}
K.~Friston, S.~Kiebel, {Predictive coding under the free-energy principle},
  Philosophical Transactions of the Royal Society B: Biological Sciences
  364~(1521) (2009) 1211--1221.
\newblock \href {https://doi.org/10.1098/rstb.2008.0300}
  {\path{doi:10.1098/rstb.2008.0300}}.

\bibitem{Prince2012}
S.~J.~D. Prince, {Computer Vision: Models, Learning and Inference}, Cambridge
  University Press, Cambridge, 2012.
\newblock \href {https://doi.org/10.1017/CBO9780511996504}
  {\path{doi:10.1017/CBO9780511996504}}.

\bibitem{DiCarlo2012}
J.~J. DiCarlo, D.~Zoccolan, N.~C. Rust, {How does the brain solve visual object
  recognition?}, Neuron 73~(3) (2012) 415--434.
\newblock \href {https://doi.org/10.1016/j.neuron.2012.01.010}
  {\path{doi:10.1016/j.neuron.2012.01.010}}.

\bibitem{Carandini2012}
M.~Carandini, D.~J. Heeger, {Normalization as a canonical neural computation},
  Nature Reviews Neuroscience 13~(1) (2012) 51--62.
\newblock \href {https://doi.org/10.1038/nrn3136} {\path{doi:10.1038/nrn3136}}.

\bibitem{Desimone1995}
R.~Desimone, J.~Duncan, {Neural Mechanisms of Selective Visual Attention},
  Annual Review of Neuroscience 18~(1) (1995) 193--222.
\newblock \href {https://doi.org/10.1146/annurev.neuro.18.1.193}
  {\path{doi:10.1146/annurev.neuro.18.1.193}}.

\bibitem{Kastner2000}
S.~Kastner, L.~G. Ungerleider, {Mechanisms of Visual Attention in the Human
  Cortex}, Annual Review of Neuroscience 23~(1) (2000) 315--341.
\newblock \href {https://doi.org/10.1146/annurev.neuro.23.1.315}
  {\path{doi:10.1146/annurev.neuro.23.1.315}}.

\bibitem{Maunsell2006}
J.~H. Maunsell, S.~Treue, {Feature-based attention in visual cortex}, Trends in
  Neurosciences 29~(6) (2006) 317--322.
\newblock \href {https://doi.org/10.1016/j.tins.2006.04.001}
  {\path{doi:10.1016/j.tins.2006.04.001}}.

\bibitem{NIPS2017_7181}
A.~Vaswani, N.~Shazeer, N.~Parmar, J.~Uszkoreit, L.~Jones, A.~N. Gomez, L.~u.
  Kaiser, I.~Polosukhin, Attention is all you need, in: I.~Guyon, U.~V.
  Luxburg, S.~Bengio, H.~Wallach, R.~Fergus, S.~Vishwanathan, R.~Garnett
  (Eds.), Advances in Neural Information Processing Systems 30, Curran
  Associates, Inc., 2017, pp. 5998--6008.

\bibitem{Liao2016}
Q.~Liao, T.~Poggio, {Bridging the Gaps Between Residual Learning, Recurrent
  Neural Networks and Visual Cortex}~(047) (2016) 1--16.
\newblock \href {http://arxiv.org/abs/1604.03640} {\path{arXiv:1604.03640}}.

\bibitem{Jastrzebski2017}
S.~Jastrz{\c{e}}bski, D.~Arpit, N.~Ballas, V.~Verma, T.~Che, Y.~Bengio,
  {Residual Connections Encourage Iterative Inference} (2017).
\newblock \href {http://arxiv.org/abs/1710.04773} {\path{arXiv:1710.04773}}.

\bibitem{Greff2019}
K.~Greff, R.~K. Srivastava, J.~Schmidhuber, {Highway and Residual Networks
  learn Unrolled Iterative Estimation}, 5th International Conference on
  Learning Representations, ICLR 2017 - Conference Track Proceedings~(2015)
  (2016) 1--14.
\newblock \href {http://arxiv.org/abs/1612.07771} {\path{arXiv:1612.07771}}.

\bibitem{Huang2017}
G.~Huang, Z.~Liu, L.~{Van Der Maaten}, K.~Q. Weinberger, {Densely connected
  convolutional networks}, Proceedings - 30th IEEE Conference on Computer
  Vision and Pattern Recognition, CVPR 2017 2017-January (2017) 2261--2269.
\newblock \href {https://doi.org/10.1109/CVPR.2017.243}
  {\path{doi:10.1109/CVPR.2017.243}}.

\bibitem{Dayan2001}
P.~Dayan, L.~F. Abbott, {Theoretical Neuroscience}, MIT Press, Cambridge, MA,
  2001.

\bibitem{Advani2017}
M.~S. Advani, A.~M. Saxe, {High-dimensional dynamics of generalization error in
  neural networks} (2017) 1--32\href {http://arxiv.org/abs/1710.03667}
  {\path{arXiv:1710.03667}}.

\bibitem{Belkin2019}
M.~Belkin, D.~Hsu, S.~Ma, S.~Mandal, {Reconciling modern machine-learning
  practice and the classical bias–variance trade-off}, Proceedings of the
  National Academy of Sciences of the United States of America 116~(32) (2019)
  15849--15854.
\newblock \href {https://doi.org/10.1073/pnas.1903070116}
  {\path{doi:10.1073/pnas.1903070116}}.

\bibitem{Nakkiran2019}
P.~Nakkiran, G.~Kaplun, Y.~Bansal, T.~Yang, B.~Barak, I.~Sutskever, {Deep
  Double Descent: Where Bigger Models and More Data Hurt} (2019).
\newblock \href {http://arxiv.org/abs/1912.02292} {\path{arXiv:1912.02292}}.

\bibitem{Rawat2017}
W.~Rawat, Z.~Wang, {Deep Convolutional Neural Networks for Image
  Classification: A Comprehensive Review}, Neural Computation 29~(9) (2017)
  2352--2449.
\newblock \href {https://doi.org/10.1162/neco_a_00990}
  {\path{doi:10.1162/neco_a_00990}}.

\bibitem{Gilbert2013}
C.~D. Gilbert, W.~Li, {Top-down influences on visual processing}, Nature
  Reviews Neuroscience 14~(5) (2013) 350--363.
\newblock \href {https://doi.org/10.1038/nrn3476} {\path{doi:10.1038/nrn3476}}.

\bibitem{Summerfield2009}
C.~Summerfield, T.~Egner, {Expectation (and attention) in visual cognition},
  Trends in Cognitive Sciences 13~(9) (2009) 403--409.
\newblock \href {https://doi.org/10.1016/j.tics.2009.06.003}
  {\path{doi:10.1016/j.tics.2009.06.003}}.

\bibitem{von1867handbuch}
H.~von Helmholtz, {Handbuch der physiologischen Optik}, Dover (English
  translation), New York, 1860/1962.

\bibitem{Weiss2002}
Y.~Weiss, E.~P. Simoncelli, E.~H. Adelson, {Motion illusions as optimal
  percepts}, Nature Neuroscience 5~(6) (2002) 598--604.
\newblock \href {https://doi.org/10.1038/nn858} {\path{doi:10.1038/nn858}}.

\bibitem{Stocker2006}
A.~A. Stocker, E.~P. Simoncelli, {Noise characteristics and prior expectations
  in human visual speed perception}, Nature Neuroscience 9~(4) (2006) 578--585.
\newblock \href {https://doi.org/10.1038/nn1669} {\path{doi:10.1038/nn1669}}.

\bibitem{Mamassian2001}
P.~Mamassian, R.~Goutcher, {Prior knowledge on the illumination position},
  Cognition 81~(1) (2001) 1--9.
\newblock \href {https://doi.org/10.1016/S0010-0277(01)00116-0}
  {\path{doi:10.1016/S0010-0277(01)00116-0}}.

\bibitem{Girshick2011}
A.~R. Girshick, M.~S. Landy, E.~P. Simoncelli, {Cardinal rules: visual
  orientation perception reflects knowledge of environmental statistics.},
  Nature neuroscience 14~(7) (2011) 926--32.
\newblock \href {https://doi.org/10.1038/nn.2831} {\path{doi:10.1038/nn.2831}}.

\bibitem{Hasson2008}
U.~Hasson, E.~Yang, I.~Vallines, D.~J. Heeger, N.~Rubin, {A Hierarchy of
  Temporal Receptive Windows in Human Cortex}, Journal of Neuroscience 28~(10)
  (2008) 2539--2550.
\newblock \href {https://doi.org/10.1523/JNEUROSCI.5487-07.2008}
  {\path{doi:10.1523/JNEUROSCI.5487-07.2008}}.

\bibitem{Murray2014}
J.~D. Murray, A.~Bernacchia, D.~J. Freedman, R.~Romo, J.~D. Wallis, X.~Cai,
  C.~Padoa-Schioppa, T.~Pasternak, H.~Seo, D.~Lee, X.-J. Wang, {A hierarchy of
  intrinsic timescales across primate cortex}, Nature Neuroscience 17~(12)
  (2014) 1661--1663.
\newblock \href {https://doi.org/10.1038/nn.3862} {\path{doi:10.1038/nn.3862}}.

\bibitem{Sutskever2014}
I.~Sutskever, O.~Vinyals, Q.~V. Le, {Sequence to sequence learning with neural
  networks}, Advances in Neural Information Processing Systems 4~(January)
  (2014) 3104--3112.
\newblock \href {http://arxiv.org/abs/1409.3215} {\path{arXiv:1409.3215}}.

\bibitem{Kalman1960}
R.~E. Kalman, {A New Approach to Linear Filtering and Prediction Problems},
  Journal of Basic Engineering 82~(1) (1960) 35--45.
\newblock \href {https://doi.org/10.1115/1.3662552}
  {\path{doi:10.1115/1.3662552}}.

\bibitem{Wolpert1995}
D.~Wolpert, Z.~Ghahramani, M.~Jordan, {An internal model for sensorimotor
  integration}, Science 269~(5232) (1995) 1880--1882.
\newblock \href {https://doi.org/10.1126/science.7569931}
  {\path{doi:10.1126/science.7569931}}.

\bibitem{Rao1997a}
R.~P.~N. Rao, D.~H. Ballard, {Dynamic model of visual recognition predicts
  neural response properties in the visual cortex}, Neural computation
  9~(November 1995) (1997) 721--763.
\newblock \href {https://doi.org/10.1162/neco.1997.9.4.721}
  {\path{doi:10.1162/neco.1997.9.4.721}}.

\bibitem{Rao2004}
R.~P.~N. Rao, {Bayesian computation in recurrent neural circuits.}, Neural
  computation 16~(1) (2004) 1--38.

\bibitem{Deneve2007}
S.~Den{\`{e}}ve, J.-R. Duhamel, A.~Pouget, {Optimal Sensorimotor Integration in
  Recurrent Cortical Networks: A Neural Implementation of Kalman Filters},
  Journal of Neuroscience 27~(21) (2007) 5744--5756.
\newblock \href {https://doi.org/10.1523/JNEUROSCI.3985-06.2007}
  {\path{doi:10.1523/JNEUROSCI.3985-06.2007}}.

\bibitem{Xivry2013}
J.-J. {Orban de Xivry}, S.~Coppe, G.~Blohm, P.~Lefevre, {Kalman Filtering
  Naturally Accounts for Visually Guided and Predictive Smooth Pursuit
  Dynamics}, Journal of Neuroscience 33~(44) (2013) 17301--17313.
\newblock \href {https://doi.org/10.1523/JNEUROSCI.2321-13.2013}
  {\path{doi:10.1523/JNEUROSCI.2321-13.2013}}.

\bibitem{Kwon2015}
O.-S. Kwon, D.~Tadin, D.~C. Knill, {Unifying account of visual motion and
  position perception}, Proceedings of the National Academy of Sciences
  112~(26) (2015) 8142--8147.
\newblock \href {https://doi.org/10.1073/pnas.1500361112}
  {\path{doi:10.1073/pnas.1500361112}}.

\bibitem{VanBergen2019}
R.~S. van Bergen, J.~F.~M. Jehee, {Probabilistic Representation in Human Visual
  Cortex Reflects Uncertainty in Serial Decisions}, The Journal of neuroscience
  : the official journal of the Society for Neuroscience 39~(41) (2019)
  8164--8176.
\newblock \href {https://doi.org/10.1523/JNEUROSCI.3212-18.2019}
  {\path{doi:10.1523/JNEUROSCI.3212-18.2019}}.

\bibitem{Graves2013}
A.~Graves, A.-R. Mohamed, G.~Hinton, {Speech recognition with deep recurrent
  neural networks}, in: 2013 IEEE International Conference on Acoustics, Speech
  and Signal Processing, no.~3, IEEE, 2013, pp. 6645--6649.
\newblock \href {https://doi.org/10.1109/ICASSP.2013.6638947}
  {\path{doi:10.1109/ICASSP.2013.6638947}}.

\bibitem{Sak2014}
H.~Sak, A.~Senior, F.~Beaufays, {Long Short-Term Memory Based Recurrent Neural
  Network Architectures for Large Vocabulary Speech Recognition} (2014).
\newblock \href {http://arxiv.org/abs/1402.1128} {\path{arXiv:1402.1128}}.

\bibitem{Bahdanau2014}
D.~Bahdanau, K.~Cho, Y.~Bengio, {Neural Machine Translation by Jointly Learning
  to Align and Translate}, 3rd International Conference on Learning
  Representations, ICLR 2015 - Conference Track Proceedings (2014).
\newblock \href {http://arxiv.org/abs/1409.0473} {\path{arXiv:1409.0473}}.

\bibitem{Cho2014}
K.~Cho, B.~van Merrienboer, C.~Gulcehre, D.~Bahdanau, F.~Bougares, H.~Schwenk,
  Y.~Bengio, {Learning Phrase Representations using RNN Encoder-Decoder for
  Statistical Machine Translation}, Journal of Clinical Microbiology 28~(4)
  (2014) 828--829.
\newblock \href {http://arxiv.org/abs/1406.1078} {\path{arXiv:1406.1078}}.

\bibitem{Ranzato2014}
M.~Ranzato, A.~Szlam, J.~Bruna, M.~Mathieu, R.~Collobert, S.~Chopra, {Video
  (language) modeling: a baseline for generative models of natural videos}
  (2014).
\newblock \href {http://arxiv.org/abs/1412.6604} {\path{arXiv:1412.6604}}.

\bibitem{Srivastava2015}
N.~Srivastava, E.~Mansimov, R.~Salakhutdinov, {Unsupervised Learning of Video
  Representations using LSTMs} (2015).
\newblock \href {http://arxiv.org/abs/1502.04681} {\path{arXiv:1502.04681}}.

\bibitem{Lotter2016}
W.~Lotter, G.~Kreiman, D.~Cox, {Deep Predictive Coding Networks for Video
  Prediction and Unsupervised Learning} (2016).
\newblock \href {http://arxiv.org/abs/1605.08104} {\path{arXiv:1605.08104}}.

\bibitem{Pouget2013}
A.~Pouget, J.~Beck, W.~J. Ma, P.~Latham, {Probabilistic brains: knowns and
  unknowns.}, Nature neuroscience 16~(9) (2013) 1170--8.
\newblock \href {https://doi.org/10.1038/nn.3495} {\path{doi:10.1038/nn.3495}}.

\bibitem{Ma2014}
W.~J. Ma, M.~Jazayeri, {Neural Coding of Uncertainty and Probability.}, Annual
  Review of Neuroscience 37 (2014) 205--220.
\newblock \href {https://doi.org/10.1146/annurev-neuro-071013-014017}
  {\path{doi:10.1146/annurev-neuro-071013-014017}}.

\bibitem{Orban2016}
G.~Orb{\'{a}}n, P.~Berkes, J.~Fiser, M.~Lengyel, {Neural Variability and
  Sampling-Based Probabilistic Representations in the Visual Cortex}, Neuron
  92~(2) (2016) 530--543.
\newblock \href {https://doi.org/10.1016/j.neuron.2016.09.038}
  {\path{doi:10.1016/j.neuron.2016.09.038}}.

\bibitem{Boerlin2013}
M.~Boerlin, C.~K. Machens, S.~Den{\`{e}}ve, {Predictive Coding of Dynamical
  Variables in Balanced Spiking Networks}, PLoS Computational Biology 9~(11)
  (2013).
\newblock \href {https://doi.org/10.1371/journal.pcbi.1003258}
  {\path{doi:10.1371/journal.pcbi.1003258}}.

\bibitem{Barrett2016}
D.~G. Barrett, S.~Den{\`{e}}ve, C.~K. Machens, {Optimal compensation for neuron
  loss}, eLife 5~(e12454) (2016) 1--36.
\newblock \href {https://doi.org/10.7554/eLife.12454}
  {\path{doi:10.7554/eLife.12454}}.

\bibitem{Schiller1976}
P.~H. Schiller, B.~L. Finlay, S.~F. Volman, {Short-term response variability of
  monkey striate neurons.}, Brain research 105~(2) (1976) 347--9.

\bibitem{Dean1981}
A.~Dean, {The variability of discharge of simple cells in the cat striate
  cortex}, Experimental Brain Research 44~(4) (1981).
\newblock \href {https://doi.org/10.1007/BF00238837}
  {\path{doi:10.1007/BF00238837}}.

\bibitem{Mainen1995}
Z.~F. Mainen, T.~J. Sejnowski, {Reliability of spike timing in neocortical
  neurons.}, Science 268~(5216) (1995) 1503--6.

\bibitem{Krizhevsky2012}
A.~Krizhevsky, I.~Sutskever, G.~E. Hinton, {ImageNet Classification with Deep
  Convolutional Neural Networks}, Advances In Neural Information Processing
  Systems (2012).
\newblock \href {http://arxiv.org/abs/1102.0183} {\path{arXiv:1102.0183}}.

\bibitem{Field1993}
D.~J. Field, A.~Hayes, R.~F. Hess, {Contour integration by the human visual
  system: evidence for a local "association field".}, Vision research 33~(2)
  (1993) 173--93.
\newblock \href {https://doi.org/10.1016/0042-6989(93)90156-q}
  {\path{doi:10.1016/0042-6989(93)90156-q}}.

\bibitem{Geisler2001}
W.~S. Geisler, J.~S. Perry, B.~J. Super, D.~P. Gallogly, {Edge co-occurrence in
  natural images predicts contour grouping performance}, Vision Research 41~(6)
  (2001) 711--724.
\newblock \href {https://doi.org/10.1016/S0042-6989(00)00277-7}
  {\path{doi:10.1016/S0042-6989(00)00277-7}}.

\bibitem{Roelfsema2006}
P.~R. Roelfsema, {Cortical algorithms for perceptual grouping}, Annual Review
  of Neuroscience 29~(1) (2006) 203--227.
\newblock \href {https://doi.org/10.1146/annurev.neuro.29.051605.112939}
  {\path{doi:10.1146/annurev.neuro.29.051605.112939}}.

\bibitem{Linsley2018b}
D.~Linsley, J.~Kim, V.~Veerabadran, C.~Windolf, T.~Serre, Learning long-range
  spatial dependencies with horizontal gated recurrent units, in: S.~Bengio,
  H.~Wallach, H.~Larochelle, K.~Grauman, N.~Cesa-Bianchi, R.~Garnett (Eds.),
  Advances in Neural Information Processing Systems 31, Curran Associates,
  Inc., 2018, pp. 152--164.

\bibitem{George2017}
D.~George, W.~Lehrach, K.~Kansky, M.~L{\'{a}}zaro-Gredilla, C.~Laan, B.~Marthi,
  X.~Lou, Z.~Meng, Y.~Liu, H.~Wang, A.~Lavin, D.~S. Phoenix, {A generative
  vision model that trains with high data efficiency and breaks text-based
  CAPTCHAs}, Science 358~(6368) (2017).
\newblock \href {https://doi.org/10.1126/science.aag2612}
  {\path{doi:10.1126/science.aag2612}}.

\bibitem{Zheng2015}
S.~Zheng, S.~Jayasumana, B.~Romera-Paredes, V.~Vineet, Z.~Su, D.~Du, C.~Huang,
  P.~H. Torr, {Conditional random fields as recurrent neural networks},
  Proceedings of the IEEE International Conference on Computer Vision 2015
  Inter (2015) 1529--1537.
\newblock \href {https://doi.org/10.1109/ICCV.2015.179}
  {\path{doi:10.1109/ICCV.2015.179}}.

\bibitem{Spoerer2017}
C.~J. Spoerer, P.~McClure, N.~Kriegeskorte, {Recurrent convolutional neural
  networks: A better model of biological object recognition}, Frontiers in
  Psychology 8~(SEP) (2017) 1--14.
\newblock \href {https://doi.org/10.3389/fpsyg.2017.01551}
  {\path{doi:10.3389/fpsyg.2017.01551}}.

\bibitem{Montobbio2019a}
N.~Montobbio, L.~Bonnasse-Gahot, G.~Citti, A.~Sarti, {KerCNNs: biologically
  inspired lateral connections for classification of corrupted images} (2019).
\newblock \href {http://arxiv.org/abs/1910.08336} {\path{arXiv:1910.08336}}.

\bibitem{Szegedy2014}
C.~Szegedy, W.~Zaremba, I.~Sutskever, J.~Bruna, D.~Erhan, I.~Goodfellow,
  R.~Fergus, {Intriguing properties of neural networks}, 2nd International
  Conference on Learning Representations, ICLR 2014 - Conference Track
  Proceedings (2014).
\newblock \href {http://arxiv.org/abs/1312.6199} {\path{arXiv:1312.6199}}.

\bibitem{Geirhos2019}
R.~Geirhos, C.~Michaelis, F.~A. Wichmann, P.~Rubisch, M.~Bethge, W.~Brendel,
  {Imagenet-trained CNNs are biased towards texture; increasing shape bias
  improves accuracy and robustness}, 7th International Conference on Learning
  Representations, ICLR 2019~(c) (2019) 1--22.
\newblock \href {http://arxiv.org/abs/1811.12231} {\path{arXiv:1811.12231}}.

\bibitem{Jacobsen2019}
J.~H. Jacobsen, J.~Behrmann, R.~Zemel, M.~Bethge, {Excessive invariance causes
  adversarial vulnerability}, 7th International Conference on Learning
  Representations, ICLR 2019 (2019).
\newblock \href {http://arxiv.org/abs/1811.00401} {\path{arXiv:1811.00401}}.

\bibitem{Neyman1933}
J.~Neyman, E.~S. Pearson, {IX. On the problem of the most efficient tests of
  statistical hypotheses}, Philosophical Transactions of the Royal Society of
  London. Series A, Containing Papers of a Mathematical or Physical Character
  231~(694-706) (1933) 289--337.
\newblock \href {https://doi.org/10.1098/rsta.1933.0009}
  {\path{doi:10.1098/rsta.1933.0009}}.

\bibitem{Engstrom2019}
A.~Ilyas, S.~Santurkar, D.~Tsipras, L.~Engstrom, B.~Tran, A.~Madry,
  {Adversarial Examples Are Not Bugs, They Are Features} (2019).
\newblock \href {http://arxiv.org/abs/1905.02175} {\path{arXiv:1905.02175}}.

\bibitem{Li2019}
Y.~Li, J.~Bradshaw, Y.~Sharma, {Are generative classifiers more robust to
  adversarial attacks?}, 36th International Conference on Machine Learning,
  ICML 2019 2019-June (2019) 6754--6783.
\newblock \href {http://arxiv.org/abs/1802.06552} {\path{arXiv:1802.06552}}.

\bibitem{Schott2018}
L.~Schott, J.~Rauber, M.~Bethge, W.~Brendel, {Towards the first adversarially
  robust neural network model on MNIST}, Iclr 3 (2018) 1--16.
\newblock \href {http://arxiv.org/abs/1805.09190} {\path{arXiv:1805.09190}}.

\bibitem{Golan2019}
T.~Golan, P.~C. Raju, N.~Kriegeskorte, {Controversial stimuli: pitting neural
  networks against each other as models of human recognition} (2019).
\newblock \href {http://arxiv.org/abs/1911.09288} {\path{arXiv:1911.09288}}.

\bibitem{Lee2003}
T.~S. Lee, D.~Mumford, {Hierarchical Bayesian inference in the visual cortex.},
  Journal of the Optical Society of America. A, Optics, image science, and
  vision 20~(7) (2003) 1434--48.

\bibitem{Wen2018}
H.~Wen, K.~Han, J.~Shi, Y.~Zhang, E.~Culurciello, Z.~Liu, {Deep Predictive
  Coding Network for Object Recognition} (2018).
\newblock \href {http://arxiv.org/abs/1802.04762} {\path{arXiv:1802.04762}}.

\bibitem{Srikumar2012}
V.~Srikumar, G.~Kundu, D.~Roth, {On amortizing inference cost for structured
  prediction}, EMNLP-CoNLL 2012 - 2012 Joint Conference on Empirical Methods in
  Natural Language Processing and Computational Natural Language Learning,
  Proceedings of the Conference~(July) (2012) 1114--1124.

\bibitem{Stuhlmuller2013}
A.~Stuhlm\"{u}ller, J.~Taylor, N.~Goodman, Learning stochastic inverses, in:
  C.~J.~C. Burges, L.~Bottou, M.~Welling, Z.~Ghahramani, K.~Q. Weinberger
  (Eds.), Advances in Neural Information Processing Systems, Vol.~26, Curran
  Associates, Inc., 2013, pp. 3048--3056.

\bibitem{Cremer2018}
C.~Cremer, X.~Li, D.~Duvenaud, {Inference suboptimality in variational
  autoencoders}, 35th International Conference on Machine Learning, ICML 2018 3
  (2018) 1749--1760.
\newblock \href {http://arxiv.org/abs/1801.03558} {\path{arXiv:1801.03558}}.

\bibitem{Marino2018}
J.~Marino, Y.~Yue, S.~Mandt, {Iterative amortized inference}, 35th
  International Conference on Machine Learning, ICML 2018 8 (2018) 5444--5462.
\newblock \href {http://arxiv.org/abs/1807.09356} {\path{arXiv:1807.09356}}.

\bibitem{Hjelm2016}
R.~D. Hjelm, K.~Cho, J.~Chung, R.~Salakhutdinov, V.~Calhoun, N.~Jojic,
  {Iterative refinement of the approximate posterior for directed belief
  networks}, Advances in Neural Information Processing Systems~(Nips 2016)
  (2016) 4698--4706.
\newblock \href {http://arxiv.org/abs/1511.06382} {\path{arXiv:1511.06382}}.

\bibitem{Krishnan2018}
R.~G. Krishnan, D.~Liang, M.~D. Hoffman, {On the challenges of learning with
  inference networks on sparse, high-dimensional data}, International
  Conference on Artificial Intelligence and Statistics, AISTATS 2018 84 (2018)
  143--151.
\newblock \href {http://arxiv.org/abs/1710.06085} {\path{arXiv:1710.06085}}.

\bibitem{Liang2015}
M.~Liang, X.~Hu, {Recurrent convolutional neural network for object
  recognition}, Proceedings of the IEEE Computer Society Conference on Computer
  Vision and Pattern Recognition 07-12-June (2015) 3367--3375.
\newblock \href {https://doi.org/10.1109/CVPR.2015.7298958}
  {\path{doi:10.1109/CVPR.2015.7298958}}.

\bibitem{Kar2019}
K.~Kar, J.~Kubilius, K.~Schmidt, E.~B. Issa, J.~J. DiCarlo, {Evidence that
  recurrent circuits are critical to the ventral stream's execution of core
  object recognition behavior}, Nature Neuroscience 22~(6) (2019) 974--983.
\newblock \href {https://doi.org/10.1038/s41593-019-0392-5}
  {\path{doi:10.1038/s41593-019-0392-5}}.

\bibitem{Nayebi2018}
A.~Nayebi, D.~Bear, J.~Kubilius, K.~Kar, S.~Ganguli, D.~Sussillo, J.~J.
  DiCarlo, D.~L. Yamins, {Task-driven convolutional recurrent models of the
  visual system}, Advances in Neural Information Processing Systems
  2018-Decem~(NeurIPS) (2018) 5290--5301.

\bibitem{Ballard1991}
D.~H. Ballard, {Animate vision}, Artificial Intelligence 48~(1) (1991) 57--86.
\newblock \href {https://doi.org/10.1016/0004-3702(91)90080-4}
  {\path{doi:10.1016/0004-3702(91)90080-4}}.

\bibitem{Findlay2003}
J.~M. Findlay, I.~D. Gilchrist, {Active Vision}, Oxford University Press, 2003.
\newblock \href {https://doi.org/10.1093/acprof:oso/9780198524793.001.0001}
  {\path{doi:10.1093/acprof:oso/9780198524793.001.0001}}.

\bibitem{Bajcsy2018}
R.~Bajcsy, Y.~Aloimonos, J.~K. Tsotsos, {Revisiting active perception},
  Autonomous Robots 42~(2) (2018) 177--196.
\newblock \href {https://doi.org/10.1007/s10514-017-9615-3}
  {\path{doi:10.1007/s10514-017-9615-3}}.

\bibitem{Russell1997}
S.~J. Russell, {Rationality and intelligence}, Artificial Intelligence 94~(1-2)
  (1997) 57--77.
\newblock \href {https://doi.org/10.1016/S0004-3702(97)00026-X}
  {\path{doi:10.1016/S0004-3702(97)00026-X}}.

\bibitem{Gershman2015}
S.~J. Gershman, E.~J. Horvitz, J.~B. Tenenbaum, {Computational rationality: A
  converging paradigm for intelligence in brains, minds, and machines}, Science
  349~(6245) (2015) 273--278.
\newblock \href {https://doi.org/10.1126/science.aac6076}
  {\path{doi:10.1126/science.aac6076}}.

\bibitem{Griffiths2015}
T.~L. Griffiths, F.~Lieder, N.~D. Goodman, {Rational Use of Cognitive
  Resources: Levels of Analysis Between the Computational and the Algorithmic},
  Topics in Cognitive Science 7~(2) (2015) 217--229.
\newblock \href {https://doi.org/10.1111/tops.12142}
  {\path{doi:10.1111/tops.12142}}.

\bibitem{Kubilius2019}
J.~Kubilius, M.~Schrimpf, K.~Kar, R.~Rajalingham, H.~Hong, N.~Majaj, E.~Issa,
  P.~Bashivan, J.~Prescott-Roy, K.~Schmidt, A.~Nayebi, D.~Bear, D.~L. Yamins,
  J.~J. DiCarlo, Brain-like object recognition with high-performing shallow
  recurrent anns, in: H.~Wallach, H.~Larochelle, A.~Beygelzimer, F.~d'Alch\'{e}
  Buc, E.~Fox, R.~Garnett (Eds.), Advances in Neural Information Processing
  Systems 32, Curran Associates, Inc., 2019, pp. 12805--12816.

\bibitem{Deng2009}
J.~Deng, W.~Dong, R.~Socher, L.-J. Li, {Kai Li}, {Li Fei-Fei}, {ImageNet: A
  large-scale hierarchical image database}, in: 2009 IEEE Conference on
  Computer Vision and Pattern Recognition, IEEE, 2009, pp. 248--255.
\newblock \href {https://doi.org/10.1109/CVPR.2009.5206848}
  {\path{doi:10.1109/CVPR.2009.5206848}}.

\bibitem{Kietzmann2019}
T.~C. Kietzmann, C.~J. Spoerer, L.~K.~A. S{\"{o}}rensen, R.~M. Cichy, O.~Hauk,
  N.~Kriegeskorte, {Recurrence is required to capture the representational
  dynamics of the human visual system}, Proceedings of the National Academy of
  Sciences 116~(43) (2019) 201905544.
\newblock \href {https://doi.org/10.1073/pnas.1905544116}
  {\path{doi:10.1073/pnas.1905544116}}.

\bibitem{Tang2018}
H.~Tang, M.~Schrimpf, W.~Lotter, C.~Moerman, A.~Paredes, J.~O. Caro,
  W.~Hardesty, D.~Cox, G.~Kreiman, {Recurrent computations for visual pattern
  completion}, Proceedings of the National Academy of Sciences of the United
  States of America 115~(35) (2018) 8835--8840.
\newblock \href {https://doi.org/10.1073/pnas.1719397115}
  {\path{doi:10.1073/pnas.1719397115}}.

\bibitem{Enns2000}
J.~T. Enns, V.~{Di Lollo}, {What's new in visual masking?}, Trends in Cognitive
  Sciences 4~(9) (2000) 345--352.
\newblock \href {https://doi.org/10.1016/S1364-6613(00)01520-5}
  {\path{doi:10.1016/S1364-6613(00)01520-5}}.

\bibitem{Fyall2017}
A.~M. Fyall, Y.~El-Shamayleh, H.~Choi, E.~Shea-Brown, A.~Pasupathy, {Dynamic
  representation of partially occluded objects in primate prefrontal and visual
  cortex}, eLife 6 (2017) 1--25.
\newblock \href {https://doi.org/10.7554/eLife.25784}
  {\path{doi:10.7554/eLife.25784}}.

\bibitem{Choi2018}
H.~Choi, A.~Pasupathy, E.~Shea-Brown, {Predictive Coding in Area V4: Dynamic
  Shape Discrimination under Partial Occlusion}, Neural Computation 30~(5)
  (2018) 1209--1257.
\newblock \href {https://doi.org/10.1162/neco_a_01072}
  {\path{doi:10.1162/neco_a_01072}}.

\bibitem{Levi2008}
D.~M. Levi, {Crowding—An essential bottleneck for object recognition: A
  mini-review}, Vision Research 48~(5) (2008) 635--654.
\newblock \href {https://doi.org/10.1016/j.visres.2007.12.009}
  {\path{doi:10.1016/j.visres.2007.12.009}}.

\bibitem{Manassi2012}
M.~Manassi, B.~Sayim, M.~H. Herzog, {Grouping, pooling, and when bigger is
  better in visual crowding}, Journal of Vision 12~(10) (2012) 13--13.
\newblock \href {https://doi.org/10.1167/12.10.13}
  {\path{doi:10.1167/12.10.13}}.

\bibitem{Manassi2016}
M.~Manassi, S.~Lonchampt, A.~Clarke, M.~H. Herzog, {What crowding can tell us
  about object representations}, Journal of Vision 16~(3) (2016) 35.
\newblock \href {https://doi.org/10.1167/16.3.35} {\path{doi:10.1167/16.3.35}}.

\bibitem{Doerig2020}
A.~Doerig, A.~Bornet, O.~Choung, M.~Herzog, {Crowding reveals fundamental
  differences in local vs. global processing in humans and machines}, Vision
  Research 167~(August 2019) (2020) 39--45.
\newblock \href {https://doi.org/10.1016/j.visres.2019.12.006}
  {\path{doi:10.1016/j.visres.2019.12.006}}.

\bibitem{Sabour2017}
S.~Sabour, N.~Frosst, G.~E. Hinton, Dynamic routing between capsules, in:
  I.~Guyon, U.~V. Luxburg, S.~Bengio, H.~Wallach, R.~Fergus, S.~Vishwanathan,
  R.~Garnett (Eds.), Advances in Neural Information Processing Systems 30,
  Curran Associates, Inc., 2017, pp. 3856--3866.

\bibitem{Sabour2018}
S.~Sabour, N.~Frosst, G.~E. Hinton, {Matrix capsules with EM routing}, Iclr
  2018~(2011) (2018) 1--12.
\newblock \href {http://arxiv.org/abs/1710.09829} {\path{arXiv:1710.09829}}.

\bibitem{ORegan2001}
J.~K. O'Regan, A.~No{\"{e}}, {A sensorimotor account of vision and visual
  consciousness}, Behavioral and Brain Sciences 24~(5) (2001) 939--973.
\newblock \href {https://doi.org/10.1017/S0140525X01000115}
  {\path{doi:10.1017/S0140525X01000115}}.

\bibitem{Buzsaki2019}
G.~Buzs{\'{a}}ki, {The Brain from Inside Out}, Oxford University Press, 2019.
\newblock \href {https://doi.org/10.1093/oso/9780190905385.001.0001}
  {\path{doi:10.1093/oso/9780190905385.001.0001}}.

\bibitem{Werbos1990}
P.~Werbos, {Backpropagation through time: what it does and how to do it},
  Proceedings of the IEEE 78~(10) (1990) 1550--1560.
\newblock \href {https://doi.org/10.1109/5.58337} {\path{doi:10.1109/5.58337}}.

\bibitem{Guerguiev2017}
J.~Guerguiev, T.~P. Lillicrap, B.~A. Richards, {Towards deep learning with
  segregated dendrites}, eLife 6 (2017) 1--37.
\newblock \href {https://doi.org/10.7554/eLife.22901}
  {\path{doi:10.7554/eLife.22901}}.

\bibitem{NIPS2018_8089}
J.~Sacramento, R.~Ponte~Costa, Y.~Bengio, W.~Senn, Dendritic cortical
  microcircuits approximate the backpropagation algorithm, in: S.~Bengio,
  H.~Wallach, H.~Larochelle, K.~Grauman, N.~Cesa-Bianchi, R.~Garnett (Eds.),
  Advances in Neural Information Processing Systems 31, Curran Associates,
  Inc., 2018, pp. 8721--8732.

\bibitem{Whittington2019}
J.~C. Whittington, R.~Bogacz, {Theories of Error Back-Propagation in the
  Brain}, Trends in Cognitive Sciences 23~(3) (2019) 235--250.
\newblock \href {https://doi.org/10.1016/j.tics.2018.12.005}
  {\path{doi:10.1016/j.tics.2018.12.005}}.

\bibitem{Lillicrap2019}
T.~P. Lillicrap, A.~Santoro, {Backpropagation through time and the brain},
  Current Opinion in Neurobiology 55 (2019) 82--89.
\newblock \href {https://doi.org/10.1016/j.conb.2019.01.011}
  {\path{doi:10.1016/j.conb.2019.01.011}}.

\bibitem{Almeida1987}
L.~Almeida, {A learning rule for asynchronous perceptrons with feedback in a
  combinatorial environment.}, Proceedings, 1st First International Conference
  on Neural Networks 2 (1987) 609--618.

\bibitem{Pineda1987}
F.~J. Pineda, {Generalization of back-propagation to recurrent neural
  networks}, Physical Review Letters 59~(19) (1987) 2229--2232.
\newblock \href {https://doi.org/10.1103/PhysRevLett.59.2229}
  {\path{doi:10.1103/PhysRevLett.59.2229}}.

\bibitem{Liao2018}
R.~Liao, Y.~Xiong, E.~Fetaya, L.~Zhang, K.~J. Yoon, X.~Pitkow, R.~Urtasun,
  R.~Zemel, {Reviving and improving recurrent back-propagation}, 35th
  International Conference on Machine Learning, ICML 2018 7 (2018) 4807--4820.
\newblock \href {http://arxiv.org/abs/1803.06396} {\path{arXiv:1803.06396}}.

\bibitem{Linsley2020}
D.~Linsley, A.~K. Ashok, L.~N. Govindarajan, R.~Liu, T.~Serre, {Stable and
  expressive recurrent vision models} (2020).
\newblock \href {http://arxiv.org/abs/2005.11362} {\path{arXiv:2005.11362}}.

\end{thebibliography}

\end{document}